%% file: paper.tex
\documentclass[sigconf]{acmart}

\usepackage{booktabs} %

\setcopyright{acmlicensed}

\copyrightyear{2017}
\acmYear{2017}
\setcopyright{acmlicensed}
\acmConference{ACSAC 2017}{December 4--8, 2017}{San Juan, PR, USA}\acmPrice{15.00}\acmDOI{10.1145/3134600.3134626}
\acmISBN{978-1-4503-5345-8/17/12}

\input{custom}

\begin{document}
\title{Breaking and Fixing Destructive Code Read Defenses}

\author{Jannik Pewny}
\affiliation{\institution{Ruhr-Universit{\"a}t Bochum}}
\email{jannik.pewny@rub.de}

\author{Philipp Koppe}
\affiliation{\institution{Ruhr-Universit{\"a}t Bochum}}
\email{philipp.koppe@rub.de}

\author{Lucas Davi}
\affiliation{\institution{Universit{\"a}t Duisburg-Essen}}
\email{lucas.davi@uni-due.de}

\author{Thorsten Holz}
\affiliation{\institution{Ruhr-Universit{\"a}t Bochum}}
\email{thorsten.holz@rub.de}

\begin{abstract}
\input{sections/abstract.tex}

\end{abstract}

\keywords{Code-Reuse Attacks and Defenses, Execute-only Memory, Destructive Code Reads}

\maketitle

\input{sections/introduction.tex}
\input{sections/background.tex}

\input{sections/adversarymodel.tex}

\input{sections/attack.tex}

\input{sections/defense.tex}

\input{sections/implementation.tex}

\input{sections/evaluation.tex}

\input{sections/discussion.tex}

\input{sections/related-work.tex}

\input{sections/conclusions.tex}
\input{sections/acknowledgements.tex}

\bibliographystyle{ACM-Reference-Format}
\bibliography{sigproc}

\input{sections/appendix.tex}

\end{document}

%% file: custom.tex
\definecolor{ForestGreen}{HTML}{228B22}
\usepackage{graphicx}
\usepackage{enumerate}
\usepackage{amssymb}
\usepackage{url}
\usepackage{tabularx}
\usepackage{amsmath}
\usepackage{listings}

\usepackage{caption}
\DeclareCaptionType{copyrightbox}
\usepackage{colortbl}
\usepackage{multicol}
\usepackage{xcolor}
\usepackage{paralist}

\usepackage{multirow}

\usepackage{subfig}

\usepackage{pifont} %
\newcommand{\xmark}{\ding{55}}

\newcommand{\tool}{\textsc{BGDX}}

\makeatletter
\lst@AddToHook{OnEmptyLine}{\vspace{-2mm}}%
\makeatother

%% file: sections/abstract.tex
Just-in-time return-oriented programming (JIT-ROP) is a powerful memory corruption attack that 
bypasses various forms of code randomization.
Execute-only memory (XOM) can potentially prevent these attacks,
but requires source code.
In contrast, destructive code reads (DCR)
provide a trade-off between security and legacy compatibility.
The common belief is that DCR provides strong protection if combined with a high-entropy code randomization.

The contribution of this paper is twofold: first, we demonstrate 
that DCR can be bypassed regardless of the underlying code randomization scheme. 
To this end, we show novel, generic attacks that infer the code layout for highly 
randomized program code.
Second, we present the design and implementation of \tool{} (\emph{Byte-Granular DCR and XOM}), a novel mitigation technique that
protects legacy binaries against code inference attacks.
\tool{} enforces memory permissions on a
byte-granular level allowing us to combine DCR and XOM
for legacy, off-the-shelf binaries.
Our evaluation shows that \tool{} is not only effective, but highly efficient, imposing only
a geometric mean performance overhead of
3.95\,\%
on SPEC.

%% file: sections/introduction.tex
\section{Introduction}
Code-reuse attacks constitute a powerful exploitation technique that is intensively used 
to subvert the control flow of modern software~\cite{Sh2007,BuRoShSa2008,fx-rop,CaFr2008}.
These attacks exploit program bugs 
to redirect the program's control flow to existing but unintended code sequences.
Defending against these attacks has become a hot topic of research. The main defense 
techniques can be roughly categorized in proposals for control-flow integrity~\cite{cfi-journal}, 
code randomization~\cite{sok-automated-software-diversity}, and code-pointer 
integrity~\cite{cpi}.
We consider code randomization schemes which
randomize the code layout of applications at different granularity, 
i.\,e., by randomizing base addresses of program modules~\cite{aslr-linux}, register assignments~\cite{heisenbyte-18}, 
the order of functions~\cite{heisenbyte-15} and basic blocks~\cite{heisenbyte-27,xifer}, or 
even the location of each instruction~\cite{ILR}. 

Recent research demonstrates that code randomization 
schemes are vulnerable to sophisticated memory disclosure attacks: just-in-time return-oriented 
programming (JIT-ROP) dynamically reads large portions of randomized code memory.
To mitigate these attacks the concept of \textit{execute-only memory}
(XOM)~\cite{xnr,readactor,hidem,readactor-plus-plus} has been proposed.
XOM denies any read access to
program memory and thereby prevents a JIT-ROP 
attack from inferring the randomized code layout.
However, deploying XOM in practice requires
to precisely
separate intermixed
code and data to avoid program crashes. 
In general, this is \textit{impossible} 
for binary executables~\cite{distinguish-code-and-data,andriesse2016dissassembly}. As such,
recent XOM schemes like XnR~\cite{xnr} and Readactor~\cite{readactor,readactor-plus-plus} require
source code leaving legacy software unprotected.
To overcome these challenges, the concept of \textit{destructive code reads}
(DCR) was
introduced~\cite{heisenbyte,near}.
The key idea is to prevent the attacker from executing code she has read before
to thwart conventional JIT-ROP attacks.

However, recent attacks against DCR~\cite{zombie-gadgets} reveal
the feasibility of
code inference attacks if the underlying code randomization scheme offers low entropy. 
Further, DCR is ineffective if program code is not freshly randomized per load,
e.\,g., an attacker can read---and thereby destroy---a program module,
then re-load it to exploit the fresh and intact version.
So far, these attacks exploit implementation pitfalls and security trade-offs
of the underlying randomization scheme.
In other words, 
DCR coupled with load-time randomization and code randomization at a finer granularity 
(e.\,g.,~\cite{ILR,heisenbyte-27,xifer,timely-rerandomization}) is believed to resist these attacks~\cite{heisenbyte,zombie-gadgets}.

\paragraph{\textbf{Contributions.}}
In this paper, we first demonstrate the contrary: DCR is ineffective even when load-time randomization 
and fine-grained code randomization are in-place. Our novel code inference attacks allow the attacker to disclose  
highly randomized code fragments \emph{without} actually reading them.
The key idea 
behind our attacks is to
perform targeted 
memory reads of only a small fraction of the memory to
deduce many other randomized code fragments.

Second, we propose \tool{}, a novel defense mechanism that can protect
commercial off-the-shelf and legacy binaries against JIT-ROP attacks,
including the aforementioned code inference attacks.
Our defense is based on the following observation: while \textit{perfect} code and data separation for legacy binaries is in general 
an undecidable problem~\cite{distinguish-code-and-data,andriesse2016dissassembly}, 
the \textit{majority} of program code and data can be reliably confirmed as either code or data. 
That is, most program code 
can be protected by XOM, whereas the remaining program parts 
can be protected by DCR. Similar to existing XOM-schemes such as Readactor~\cite{readactor}, we 
leverage Intel's recent virtualization extension called EPT (extended page tables) to manage 
memory access rights. However, EPT enforces memory access rights on the 
granularity of 4KB pages. As a consequence, an entire page needs to be protected with DCR 
if only a couple of bytes are not distinguishable.
To reduce the DCR attack surface to a minimum,
we developed a new \emph{byte-granular} memory permission system on top of EPT,
which supports read-only, execute-only and DCR protection on a single page.
As we will show, \tool\ significantly limits code inference attacks since
most of the code memory is not readable, i.\,e.,
memory probing will almost certainly
crash the program.

\smallskip \noindent
In summary, our main contributions are as follows:

\begin{itemize}
\item We question the common belief that DCR~\cite{heisenbyte,near} is an
effective defense against code-reuse attacks. In contrast to previous DCR
attacks~\cite{zombie-gadgets}, we
demonstrate that DCR coupled with \textit{strong} code randomization is insecure,
by implementing novel, generic code inference attacks
for recent versions of Firefox (32~bit) and Internet Explorer (64~bit).
\item We show how to conceptually bypass DCR regardless of the
underlying code randomization scheme by combining a novel variant of
code inference with whole-function reuse. We provide a proof-of-concept
exploit against a recent version of Firefox (32~bit) and gathered data
suggesting that this attack is feasible against
Internet Explorer (64~bit).
\item We propose a novel defensive scheme called \emph{Byte-Granular DCR and XOM} (\tool) which protects legacy binaries
against state-of-the-art code inference attacks
based on a novel byte-granular permission system.
\item We implement a prototype of \tool{}
and test its effectiveness with our attacks.
\tool{} imposes a geometric mean performance overhead of only 
3.95\,\%
on the SPEC 2006 CPU
benchmarks and
7.9\,\%
on common browser benchmarks.
\end{itemize}

%% file: sections/background.tex
\section{Problem Setting}
\label{background}

\begin{figure}[t]
	\centering
		\includegraphics[width=\linewidth]{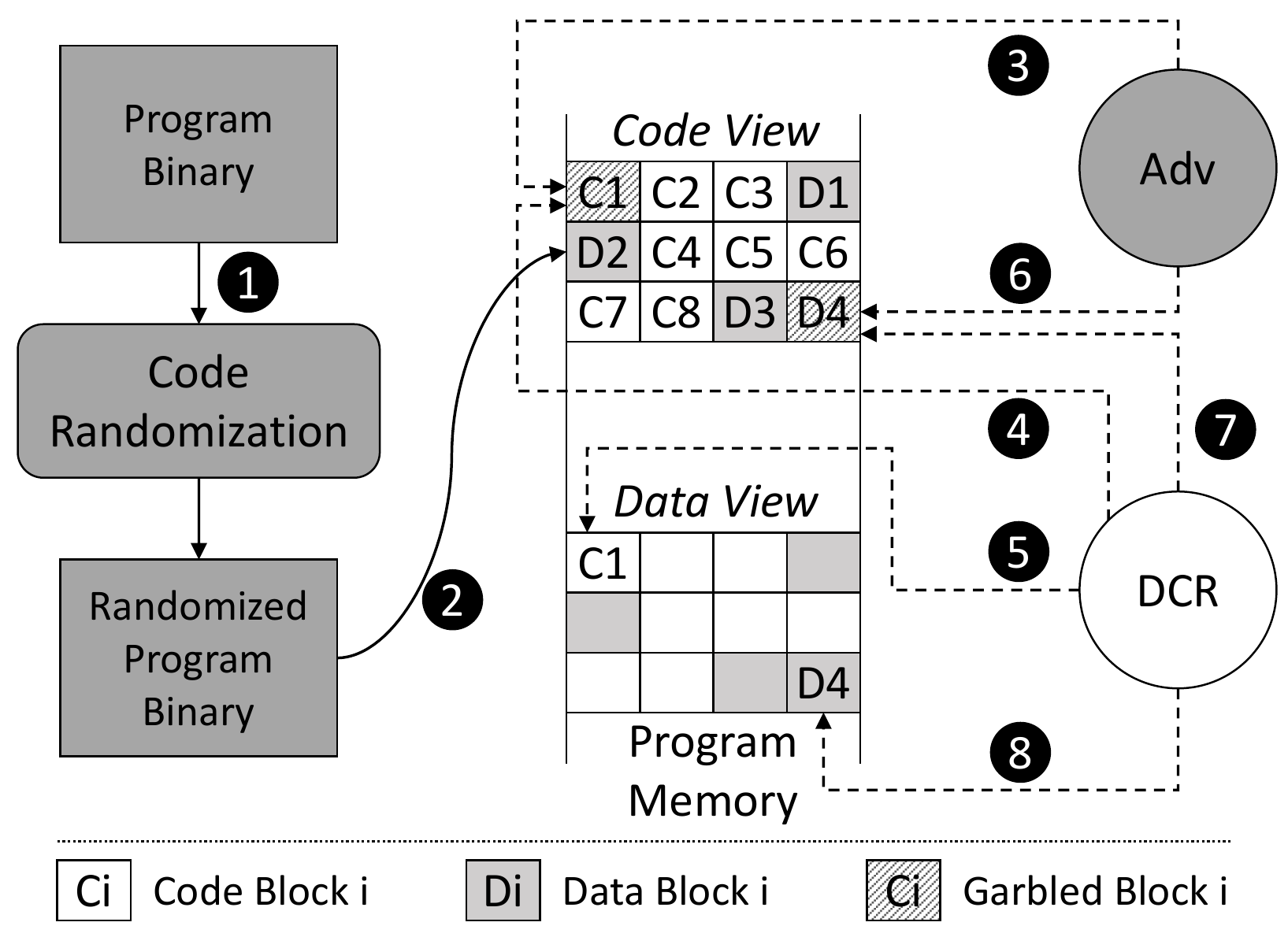}
	\caption{Overview of destructive code reads (DCR).}
	\label{fig:DCR-ROP}
\end{figure}

Destructive code reads (DCR) prevent the execution of code that has been 
read previously~\cite{heisenbyte,near}
to thwart JIT-ROP attacks~\cite{jit-rop}.
An overview on DCR is depicted in Figure~\ref{fig:DCR-ROP}: first, binary instrumentation techniques 
are leveraged to randomize the program binary~(\ding{182})
before it is loaded into program memory~\ding{183}.
Hence, an attacker cannot predict the location of 
code she aims to utilize in a code-reuse attack, and needs to 
resort to a JIT-ROP attack.

DCR distinguishes between a code and a data view on a program's code section.
As a result, DCR does not need to separate data embedded in the code section thereby not requiring 
the program's source code.
Whenever DCR observes a data fetch in the code section, DCR garbles the fetched byte in the code view with an invalid opcode, e.\,g., 
in Figure~\ref{fig:DCR-ROP}, the attacker reads code block C1 in~\ding{184}, and DCR immediately 
garbles the block in~\ding{185}. Hence, though the attacker disclosed the randomized content of 
C1, she cannot re-use this code as it has been garbled.

To preserve the program's functionality, DCR needs to support legitimate reads on data embedded in code memory. 
To do so, the values garbled in the code view are preserved in the data view in~\ding{186}. 
Thus, for subsequent read accesses, DCR returns the original bytes rather than the invalid opcodes. 
E.\,g., the read access in Figure~\ref{fig:DCR-ROP}, to the legitimate embedded data block D4 
in~\ding{187} results in garbling the code block in~\ding{188},
but later reading attempts of D4 are correctly 
handled via the stored value in~\ding{189}.

\subsection{Existing DCR Attacks}
Snow et al.~\cite{zombie-gadgets} recently introduced first attacks 
against DCR.
However, as we will argue in the following, these attacks have limited impact as they only 
exploit implementation pitfalls and mostly violate the assumptions of DCR implementations.
In particular, the DCR implementation Heisenbyte (Assumptions, p.\,3)~\cite{heisenbyte} states:
\begin{quote}
``\textit{Load-time fine-grained ASLR}: [...] we require code layouts to be randomized at a fine granularity so that the
registers~\cite{heisenbyte-18} used and instruction locations within a
function~\cite{heisenbyte-15} or basic block~\cite{heisenbyte-27} are different."
\end{quote}

Note that altered instruction locations within a function \textit{or} basic
block are required.
Thus, a scheme providing altered instruction locations within a function,
but not within basic blocks, is sufficient to fulfill this requirement.
However, Snow et al.~\cite{zombie-gadgets} 
assume a weak code randomization scheme~\cite{heisenbyte-18}
that (1)~performs in-place transformations,
i.\,e., does \textit{not}
alter offsets between the module base and individual functions
or basic blocks
and (2)~does not perform code randomization every time a module is loaded.
Given these relaxed assumptions, the following attacks are feasible: 

\begin{enumerate}
	\item Code cloning abuses a just-in-time compiler to create identical
copies of a
code region. 
	Now, the attacker can disclose gadgets of one copy, but execute
    those of the second copy.
	\item Code reloading
allows
to first disclose the code of a shared library
and, upon reload, invoke gadgets in the fresh copy.
	\item Disclosing a few bytes allows to reveal the code randomization of
close-by gadgets.
\end{enumerate}

With proper load-time randomization 
in place,
code cloning and code reloading
are prevented since the code 
is newly randomized on load.
The remaining code inference attack exploits
the narrow scope of the performed code transformations.
As mentioned, Heisenbyte requires transformations with a wider scope~\cite{heisenbyte-15, heisenbyte-27}.

In other words, DCR coupled with a code randomization scheme that offers high entropy and randomizes 
application code per load (e.\,g.,~\cite{ILR,heisenbyte-27,xifer}) resists
the attacks described by Snow et al.~\cite{zombie-gadgets}.
However, in this paper,
we go beyond attacking 
potential DCR implementation pitfalls, but explore generic code inference attacks that work 
in the presence of highly randomized code.

%% file: sections/adversarymodel.tex
\subsection{Adversary Model}
\label{adversary-model}

Our adversary model is in line with previous code reuse research,
like JIT-ROP~\cite{jit-rop}, DCR defense~\cite{heisenbyte,near} and load-time code randomization~\cite{heisenbyte-27,xifer}.
Our defense intends to prevent code-reuse attacks,
specifically JIT-ROP and code inference attacks.
I.\,e., data-only attacks~\cite{data-only,dop}
or using side-channels to deduce memory content~\cite{heisenbyte-20} is beyond the scope of this work.

\vspace{-2mm}

\paragraph{\textbf{Known binary.}}
The attacker has a copy of the original binary (before the randomization is applied).
We assume that this binary is not heavily obfuscated, such that functions,
the call graph, the control-flow graphs, and gadget locations can be derived reliably.
\vspace{-2mm}
\paragraph{\textbf{Memory disclosure vulnerability.}}
The binary has a repeatedly exploitable, byte-granular memory disclosure vulnerability
which enables the attacker to read from any address in readable memory.
\vspace{-2mm}
\paragraph{\textbf{Control-flow hijacking.}}
The binary suffers from a memory corruption vulnerability allowing the attacker to
alter writable memory containing
control data such as function pointers, 
virtual function table pointers, and return addresses.
\vspace{-2mm}
\paragraph{\textbf{Execution context.}}
The attacker has access to an execution engine to
repeatedly
issue memory reads 
and analyze disclosed memory
regions.
Modern applications, e.\,g., web browsers and PDF viewers,
often provide such execution contexts via scripting languages like JavaScript or Flash.
\vspace{-2mm}
\paragraph{\textbf{Destructive code reads.}}
Code which has been read
causes the program to crash
when it is executed at a later point
in time
due to the presence of a destructive code read defense (see Section~\ref{background}).
\vspace{-2mm}
\paragraph{\textbf{Randomization with high entropy.}}
A fine-grained load-time code randomization is applied.
Details of the randomization are given separately for our attacks presented in
Section~\ref{approach-level-1} and~\ref{approach-level-2}.
\vspace{-2mm}

\paragraph{\textbf{Conditions for successful exploitation.}}
We deem an attack successful if the attacker can execute code of her choice in
the context of the attacked process.
Since modern systems enforce data execution prevention (DEP), the attacker has to leverage code-reuse attack techniques.
To successfully execute a code-reuse attack in the presence of destructive code reads,
the attacker must take care \emph{not} to
(1)~read any byte of the gadgets invoked during the code-reuse attack, 
(2)~execute any byte that the process has previously read, and
(3)~read any byte that the process invokes during execution.

%% file: sections/attack.tex
\section{High-Level Attack Strategy}
\label{high-level-attack}

\captionsetup[subfloat]{labelformat=empty} %
\begin{figure}%
\captionsetup[subfigure]{justification=centering}
\centering
\subfloat[Original Program]{\includegraphics[width=0.235\textwidth]{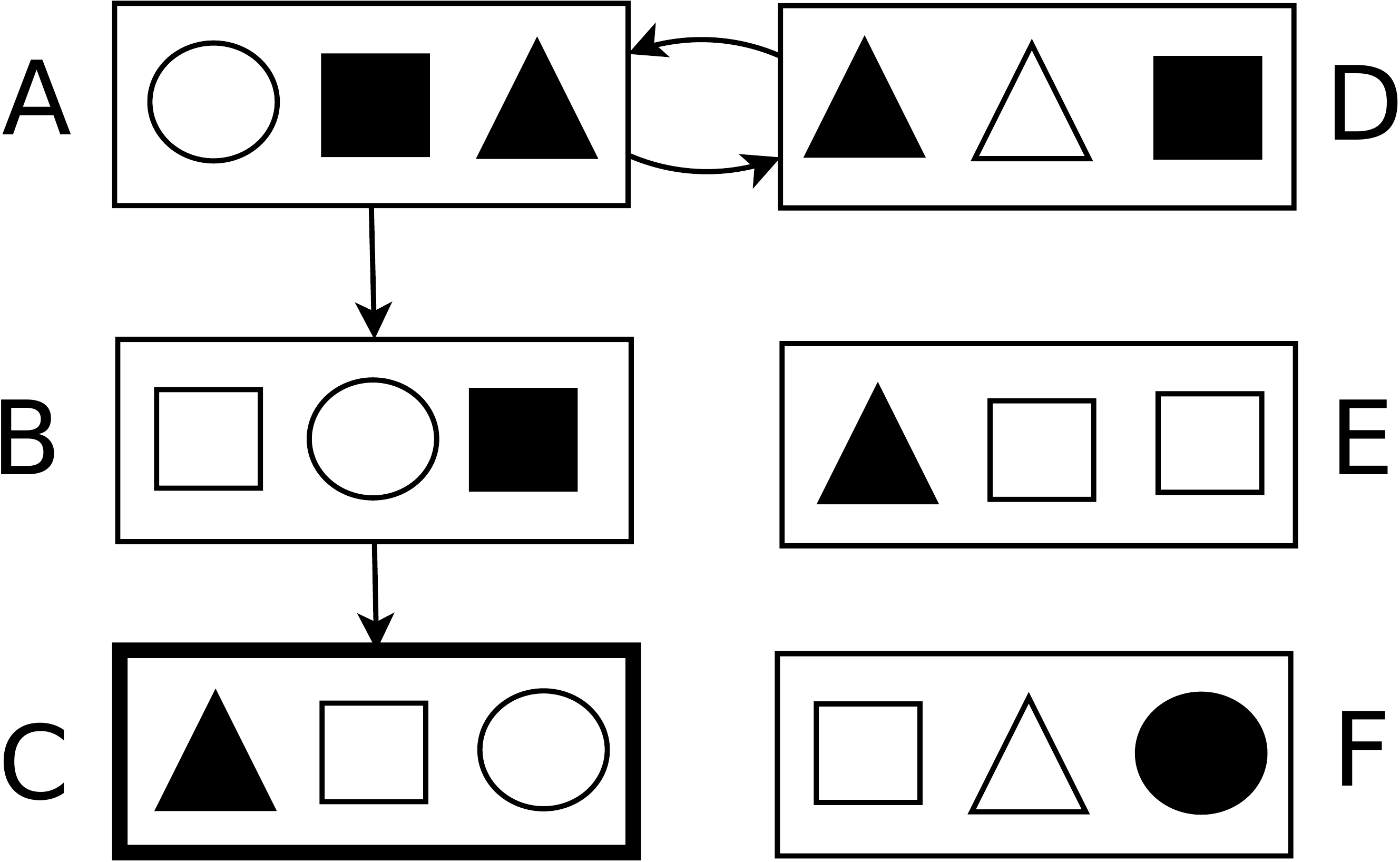}}~~~~
\subfloat[Shuffled Program]{\includegraphics[width=0.235\textwidth]{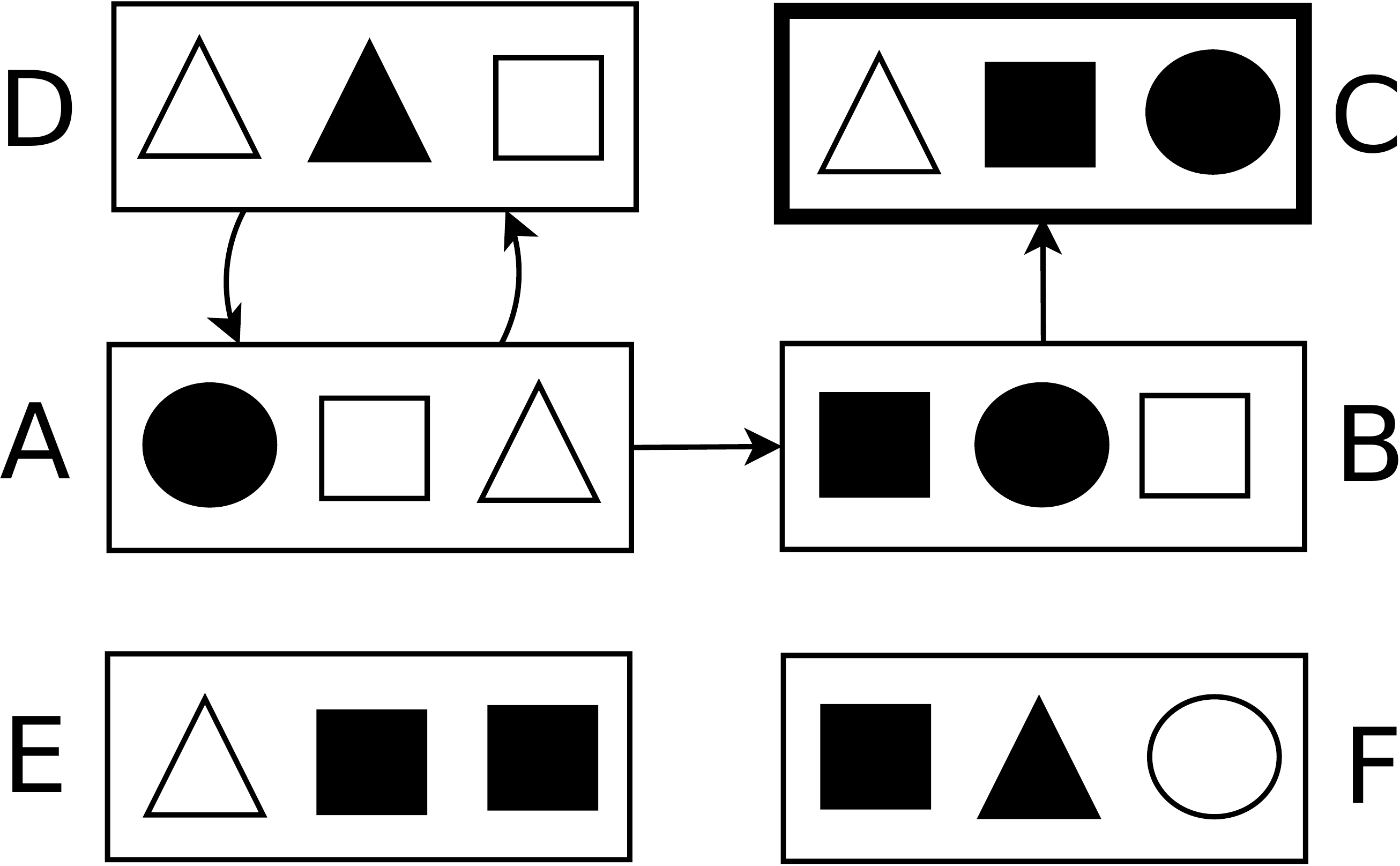}}\\
\vspace{-3mm}
~~$\Longrightarrow$\\
\vspace{-1mm}
\hrulefill\\
\vspace{-2mm}
\subfloat[\textbf{Phase 1:} Finding code anchors by probing randomly]{\includegraphics[width=0.235\textwidth]{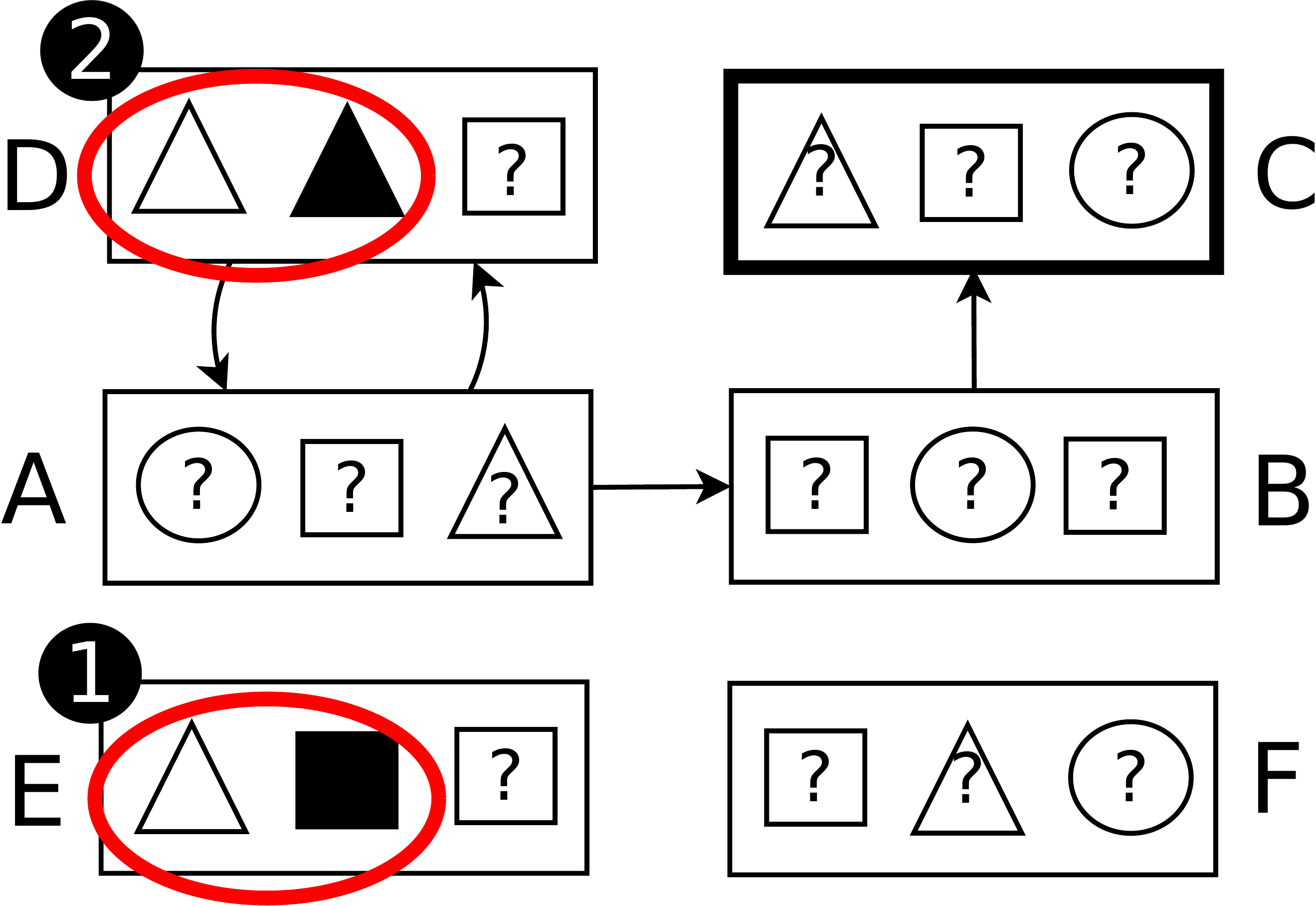}}~~~~
\subfloat[\textbf{Phase 2:} Path-guided code re-discovery to locate gadget]{\includegraphics[width=0.235\textwidth]{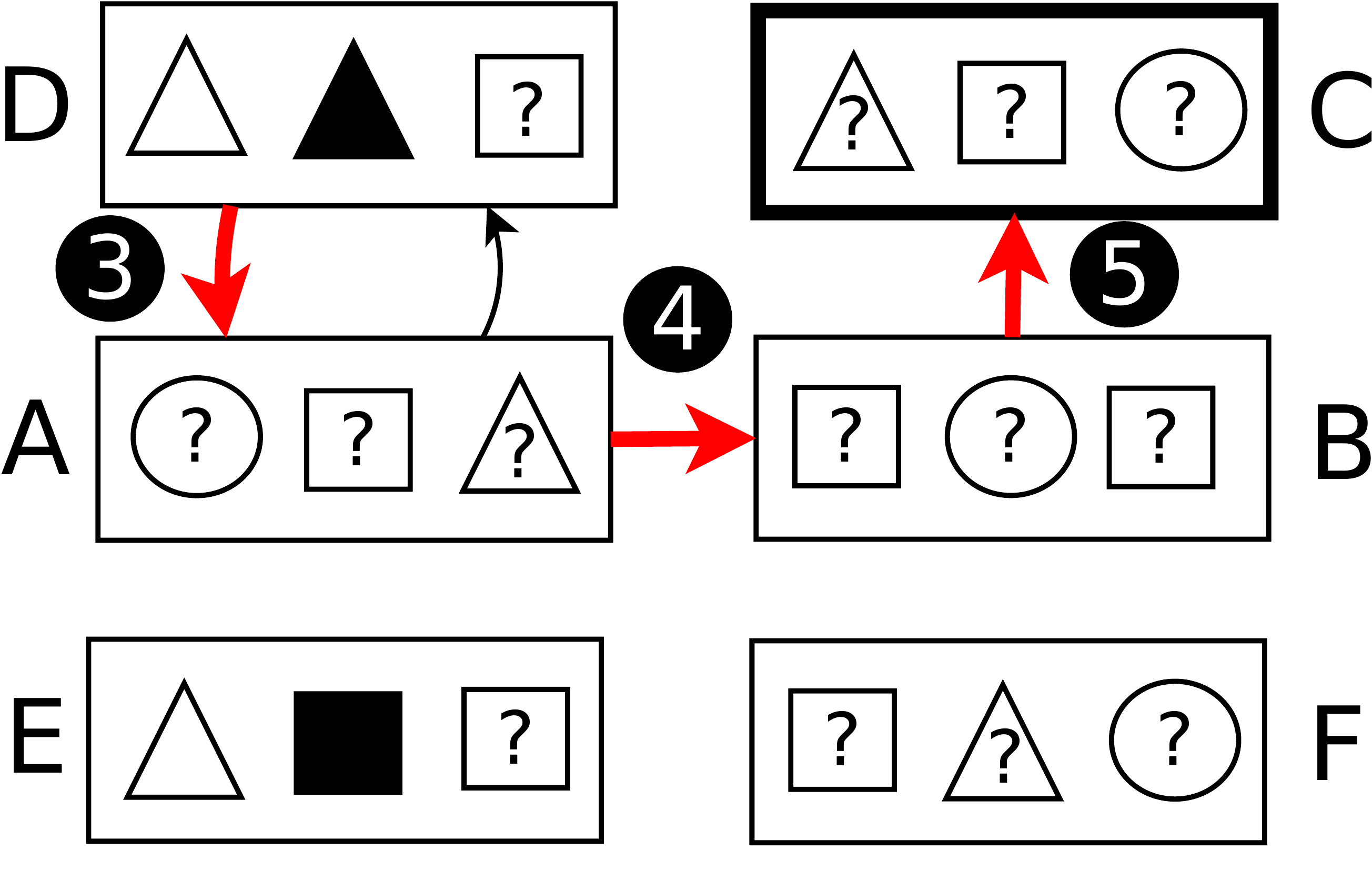}}\\
\subfloat[\textbf{Phase 3:} Gadget setup by deducing gaps through data-flow]{\includegraphics[width=0.235\textwidth]{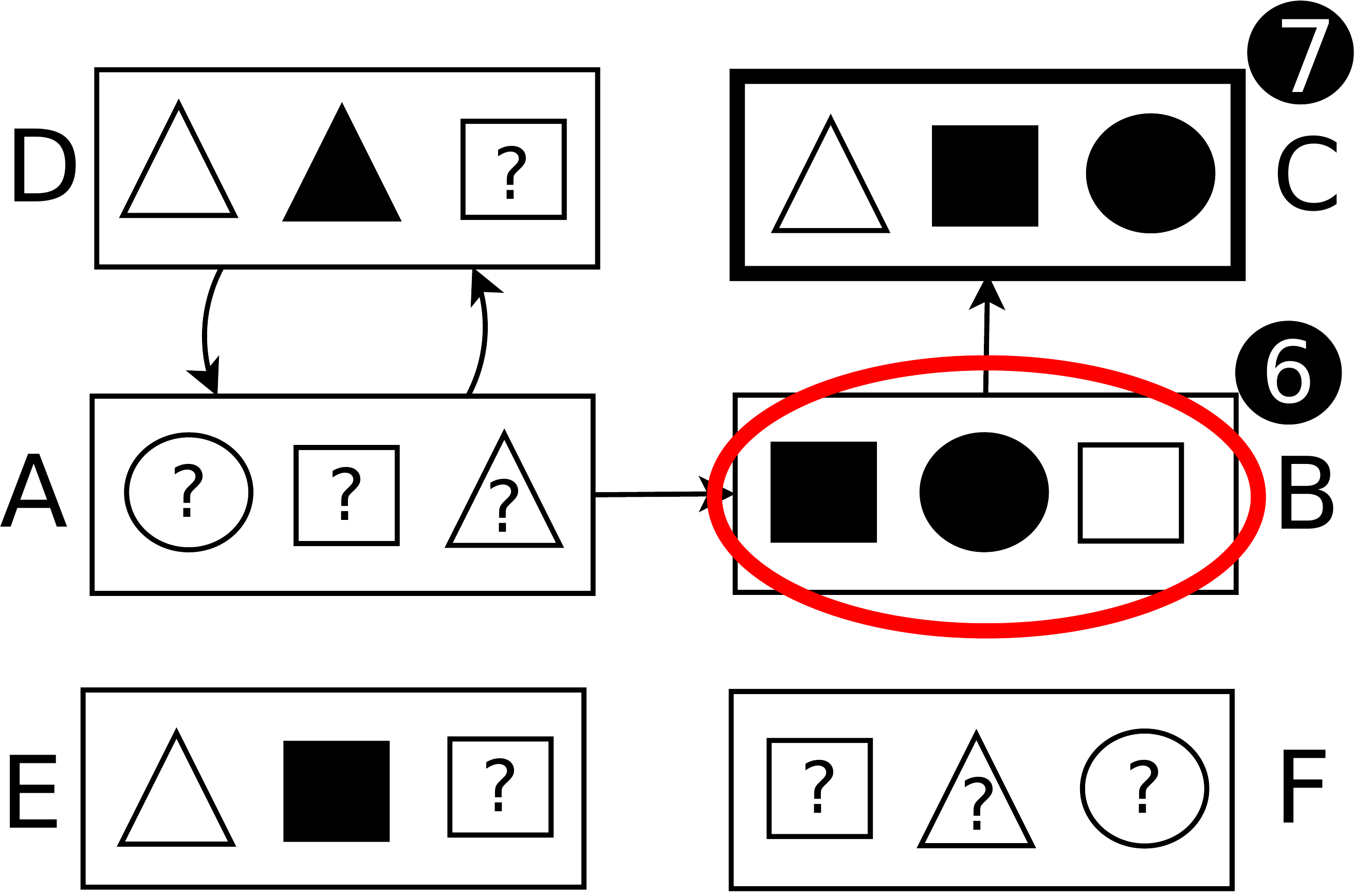}}~~~~
\subfloat[\textbf{Phase 4:} Found gadget for ROP chain without reading it]{\includegraphics[width=0.235\textwidth]{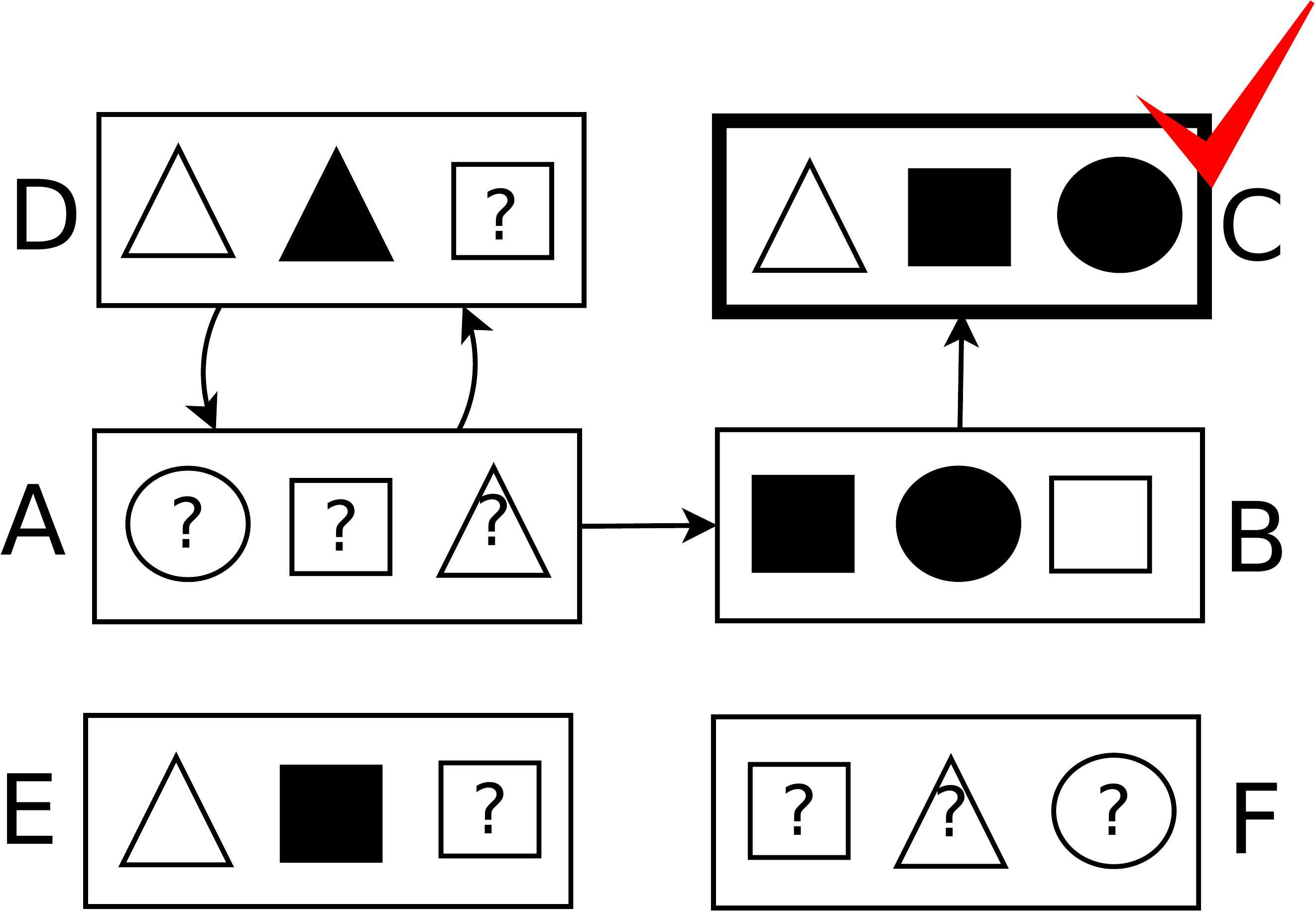}}\\

\caption{Basic workflow of our DCR bypass.}
\label{figure-attack-high-level}
\end{figure}

In the following, we provide a high-level overview of our two novel attacks
to bypass DCR schemes. 
This demonstrates general limitations of DCR and we argue
that this concept provides no viable defense in the long run.

Our first attack assumes a code randomization based on shuffled basic blocks
and registers.
This allows to deduce a program location from leaked
content at that location.
Thus, only an initial leak of the application's image base is required.
Considering the code randomization, this supposedly should not aid the attacker.
In contrast, the second attack overcomes stronger code randomization,
incorporating the full range of instruction transformations and
shuffled parameter sequences,
but requires leaking a sufficient number of code pointers.

Recall that DCR assumes that code has to be read in order
to be used as a gadget.
To bypass DCR,
the attacker has to find suitable gadgets
while
reading only code which is not executed---especially
the gadgets themselves must be missed purposefully. 

To prepare the attack, the attacker analyzes the original binary.
In particular, she extracts the call graph, the control-flow graph and
gadgets.
The attacker also extracts occurring bit patterns
which will not be modified by later code randomization.
E.\,g., when the registers of an instruction are permuted,
only the bits encoding the register will change,
while the remaining bits will stay intact.
During the attack,
these bit patterns provide important hints to help deduce which basic
block is mapped to the probed location.

The actual attack has four distinct phases: (1)~finding code anchors, 
(2)~path-guided code re-discovery, (3)~gadget setup, and (4)~executing the code-reuse attack.

Figure~\ref{figure-attack-high-level}
illustrates these phases.
For reference, the top-left subfigure shows six basic blocks (\textbf{A-F}) of the
original program memory.
Each basic block contains different byte sequences.
With the shape of a byte sequence
($\Box$, $\triangle$, $\bigcirc$), we depict parts
which are left intact by
the code randomization,
i.\,e., the shape will stay recognizable.
Note that a shape can be empty or filled to represent different
manifestations of code randomization, e.\,g., changed register-encoding
bits due to randomized register allocation.
Basic block~\textbf{C} is highlighted since it
contains a gadget the attacker aims to use.

The top-right subfigure shows the program memory after code randomization:
the locations of the basic blocks are randomized,
and the fill-state of the shapes is inverted to denote a different
manifestation of the code randomization.
When beginning the attack,
the shapes are shown with a question mark (\textbf{?}),
because the fill-state of a shape is initially unknown to the attacker.

The first phase of our attack is concerned with finding
a partial mapping between
code fragments in the randomized
process memory and the corresponding location in the original binary.
We call such a mapping a \emph{code anchor}.
To find a code anchor, the attacker randomly probes the memory.
E.\,g., the first probe in Figure~\ref{figure-attack-high-level} reveals a shape sequence which the attacker knows to
occur multiple times in the program memory~\ding{182}.
Thus, it fails to uniquely identify a location.
In contrast, the second probe~\ding{183} succeeds: the revealed sequence
only occurs in basic block \textbf{D}.
The unique shape sequence
tells the attacker that the probed location corresponds to one
exact location in the original binary, i.\,e., she just found a code anchor.
In Section~\ref{finding-code-anchors},
we provide empirical evidence showing that code anchors can be found with
high probability with only a few probes.

In the second phase, we use these code anchors to
perform a \textit{path-guided code re-discovery}
to deduce the locations of code which the attacker aims to leverage in her code-reuse attack (i.\,e., basic block \textbf{C} in Figure~\ref{figure-attack-high-level}).
We proceed in this fashion to minimize the amount of memory we have to read,
as this minimizes the chance of DCR terminating the process,
while simultaneously maximizing the chance to leave gadgets intact for later
use.
In Figure~\ref{figure-attack-high-level}, 
the attacker traverses the path
\textbf{D} $\rightarrow$ \textbf{A} $\rightarrow$ \textbf{B} $\rightarrow$ \textbf{C}
known from the original
binary by reading the basic blocks' terminating instructions (\ding{184},\ding{185})
to deduce the location of the gadget in basic block \textbf{C} (\ding{186}).
Note that this algorithm is target-oriented, i.\,e., it searches for one specific
gadget. This fundamentally distinguishes it from the dynamic discovery
of code pages in the original JIT-ROP attack~\cite{jit-rop}.

Depending
on the randomization scheme, the attacker may know \textit{which} semantic code
is present at a certain location, but may not know the \textit{syntactic
details} of that code fragment.  For example, randomized registers (see
Section~\ref{approach-level-1}) or parameter sequences (see
Section~\ref{approach-level-2}) may demand
a \textit{gadget setup} to find out which
registers or which parameter sequences are used, respectively---all
without reading the actual code fragment.
In step~\ding{187} the attacker reads basic block \textbf{B},
which precedes the targeted basic block \textbf{C}.
Given that code randomization must preserve the continuous flow of data of the original
program, the randomization of basic block \textbf{C} usually depends on the
randomization used in basic block \textbf{B}.
Thus, by comparing the randomized basic block to the basic block
from the original binary, the attacker
can trace
the data flow, and can thereby deduce the randomization used in
the targeted basic block~(\ding{188}).
Note that the targeted basic block \textbf{C} has not been read and is therefore
not garbled due to a destructive code read.

Once the gadget locations and all information to invoke them are
known, executing the ROP chain in phase 4 is straightforward.

\subsection{Minimizing Probed Memory}

In presence of DCR, every probed byte poses the risk to terminate the process
when it is executed later on.
Thus, we face the challenge
to minimize the amount of memory we have to read.
To tackle this problem, we use what we call a \emph{code anchor}.
A code anchor
maps the location of a code fragment ($address_{orig}$) in the original binary
to the corresponding location in the randomized process ($address_{rand}$).
Furthermore, we leverage the control flow of the original binary.
Even in a randomized process, a basic block's successors
can usually be determined reliably by analyzing its
terminator instruction,
which allows us to follow the control flow in a forward direction.

\begin{figure*}
\centering
\begin{minipage}{0.90\textwidth}
\lstset{language=C,
        basicstyle=\footnotesize,
        keywordstyle=\color{blue}\ttfamily,
        stringstyle=\color{red}\ttfamily,
        commentstyle=\color{ForestGreen}\ttfamily,
        morecomment=[l][\color{magenta}]{\#},
        numbers = left,
        numberblanklines=false,
        frame=lrtb,
        caption={Path-guided code re-discovery.},
        label={listing-path-guided},
        captionpos=t,
        mathescape
}
\begin{lstlisting}
Input:  Code anchor: $address_{orig} \mapsto address_{rand}$
        Target code: $t$, address in the original binary
        CFG of the original binary: $cfg$
Output: Address of the target code in the randomized process
// Use CFG to find a path between the target code and the code anchor.
 $p$ := find_path($cfg$, $address_{orig}$, $t$))
// Starting at the code anchor in the randomized process...
 $cur$ := $address_{rand}$
  while($p$ != $\emptyset$):
    // disassemble enough of the current basic block to determine its successors.
     $succs$ := get_successors($cur$)
    // Follow the equivalent edge, according to the predetermined path.
    // E.\,g., if the path followed a true-edge, follow the true-edge here, too.
     $cur$ := follow_equivalent_edge($succs$, $p$.pop())
// At the path's end, we reached the randomized basic block containing the target code.
 return $cur$
\end{lstlisting}
\end{minipage}
\end{figure*}

Listing~\ref{listing-path-guided} shows the basic approach of the path-guided
code re-discovery: first, we discover a path $p$ from the target code $t$ (e.\,g., basic block \textbf{C} in Figure~\ref{figure-attack-high-level}) to an
address in the original binary, for which we know the location in the randomized
binary (line~7).  For this, we use standard graph algorithms
on the original binary's control-flow graph and call graph.
Starting at the first basic block of the path $p$, we disassemble its
terminator
to determine the successors in the control flow
(line~14).
With those, we follow the equivalent edge in the
control flow of the randomized process (line~18).
I.\,e., if
the path $p$ in the original binary follows the \textit{true} edge of a branch,
we follow the \textit{true} edge as well.  We repeat this, node by node, until
we reach the end of the path.  At this point, we have determined the address of the
target code's basic block in the randomized process \emph{without} reading it.

\subsection{Testbed for the Attacks}
We assume
Windows~7 as the target operating
system.
There are no substantial reasons as to why our attacks or
defense could not be applied to Linux; we simply chose to
match the platform used for the Heisenbyte system.
We target a recent version of the Firefox browser v44.0.2 and Internet Explorer 11,
because browsers are a popular choice for memory-corruption attacks.
In our attacks, we leverage their main C++ libraries,
\texttt{xul.dll} respectively \texttt{mshtml.dll},
because they offer a large code base and also feature
a large number of virtual functions (i.\,e., addresses that are prone to be
leaked).
Since both libraries come with debug information,
we can use the IDA Pro disassembler~\cite{idapro}
to obtain a reliable disassembly,
including
the control-flow graphs,
the call graph, boundaries of basic blocks, and gadget locations.

Furthermore, we backported the vulnerability CVE-2014-1513~\cite{cve-2014-1513}
which skipped certain checks regarding the length and the current state of a JavaScript array.
With some engineering, this vulnerability provides an arbitrary read/write primitive,
which not only allows us to probe memory, but also to divert the control flow.

Since there is no DCR system with code randomization
publicly available\footnote{
While the authors of NEAR~\cite{near} published its source code,
it does not include a code randomization scheme.
}, we simulate its effects.
Note that this simulated testbed is used only for evaluating our attacks,
but is \textit{not} used for our defense scheme proposed in Section~\ref{defense}.
\vspace{-2mm}
\paragraph{\textbf{Simulating destructive code reads.}}
To evaluate our attacks, we
have to determine whether an attack succeeded.
Thus, we use
DynamoRIO~\cite{dynamorio} to dynamically instrument the browsers so that we can
compare executed memory locations with the ones read
during the attack:
if they are mutually exclusive, the attackers ROP chain executes successfully.
\vspace{-2mm}
\paragraph{\textbf{Simulating randomization schemes.}}
Given that implementing a code randomization scheme is not a trivial task,
we opted to simulate its effects
using a custom
oracle to answer the attacker's leaking attempts.
In particular, we perform transformations like shuffling the basic block
sequence, but return random data, e.\,g., for register-encoding bits.
This actually overestimates the effects of the code randomization,
as the attacker receives data which is less useful for deducing other code fragments.

\section{DCR Bypass of Randomized Registers and Basic Block Locations}
\label{approach-level-1}

For our first attack, we assume a load-time randomization scheme in which
the register assignment is permutated, and
each basic block is placed at a random location.
Since this alters the instruction locations within a function,
this scenario fully satisfies the requirements stated by Heisenbyte.
It also resembles a stronger randomization scheme
than the instances actually deployed in current commodity
operating systems and also stronger than the schemes considered in previously
published attacks~\cite{zombie-gadgets}.

This attack essentially probes the application's
executable memory to deduce program locations from the probed memory.
It assumes that the image base can be leaked.
Note that this supposedly does not aid the attacker,
since no gadget in the process should be in the same location due to the
code randomization.
Unlike our second attack, which requires leaked function pointers,
this attack targets a wider range of applications,
namely those, which do not stem from C\verb-++- or otherwise feature a lot
of function pointers.

\vspace{-2mm} 
\paragraph{\textbf{Phase 1: Finding Code Anchors}}
\label{finding-code-anchors}
First, we need to find code anchors.
To do so,
the attacker probes a few consecutive bytes from a random location in code memory.
Recall that she does not know,
which basic block the code randomization mapped to the probed location.
However, she does know the content of the basic blocks from the original binary.
Thus, every disclosed byte can be seen as a constraint
on the possible basic blocks at that location.
In other words,
the original basic blocks form a \textit{template}
the leaked bytes have to match.
Due to register assignment randomization,
the bits encoding a register in an instruction are initially unknown to the attacker.
Since the location of these register encoding bits can be extracted from the original
binary, we can introduce \textit{gaps} in the templates for those bits.
Consider Step~\ding{183} in Figure~\ref{figure-attack-high-level}:
the attacker discloses the byte sequence ($\triangle$, \scalebox{1.3}{$\blacktriangle$}).
Only the template (\hspace{-1mm}
{\includegraphics[width=0.015\textwidth]{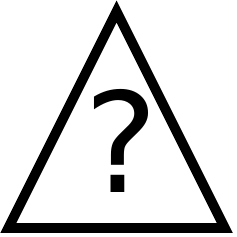}},
\includegraphics[width=0.015\textwidth]{img/triangle-question-crop}),
where the \textbf{?} denotes the gaps due to register encoding bits,
from basic block \textbf{D} matches this constraint.

\label{evaluation-finding-code-anchors}

\begin{figure}
\centering
  \includegraphics[width=0.49\textwidth]{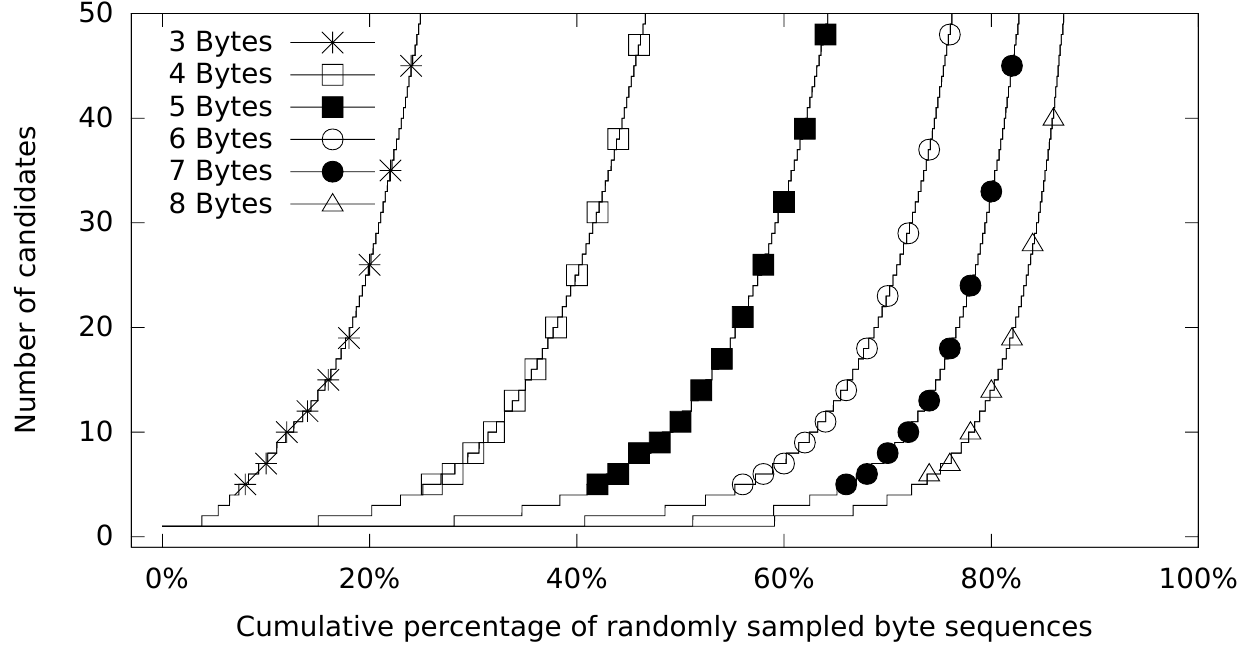}
  \caption{Number of candidates in \texttt{xul.dll} for a randomly sampled byte sequence.}
  \label{figure-byte-sequence}
\end{figure}

To determine how discriminative a leaked sequence of bytes
is, we sampled 50,000 random byte sequences of three to eight bytes from
\texttt{xul.dll} and counted how often this sequence occurs in the library
(see Figure~\ref{figure-byte-sequence}).
One also has to take
into account that the leaked sequence might overlap a basic block boundary.
Naturally, an $m$-byte sequence
can appear $1
+ n - m$ times in an $n$-byte basic block, as the last $m-1$ start positions
have to be excluded.
Considering the basic block sizes in the program
(see Appendix~\ref{appendix-bb-sizes}),
one can
calculate
the chances for a leaked $m$-byte sequence not to overlap basic block boundaries
(see Table~\ref{table-fits-chance}).
Since the considered browsers do not execute large
portions of their code (see Appendix~\ref{evaluation-code-usage}),
there is only a marginal risk for the attacker
to crash the application
by reading code the application executes.

\vspace{-2mm}
\paragraph{\textbf{Example for a six-byte sequence.}}
A six-byte sequence is unique with a chance of roughly 40\,\% (see Figure~\ref{figure-byte-sequence}).
With a chance of 75\,\%, it does not overlap a basic block boundary (see Table~\ref{table-fits-chance}).
Consequently,
there is a $30\,\%$ chance ($40\,\% \cdot 75\,\%$)
to know the location of the leaked sequence in the original program.
Thus, one can expect
to find a code anchor after leaking only four six-byte sequences ($\left\lceil{\frac{1}{30\,\%}}\right\rceil = 4$).

\begin{table}
\centering
  \caption{Chance for a randomly sampled $n$-byte sequence to not overlap basic block boundaries in \texttt{xul.dll}.}
\vspace{-2mm}

\setlength\tabcolsep{4pt}
    \begin{tabular}{c|cccccc}
\textbf{n} & 4 & 5 & 6 & 7 & 8 \\
\hline
\textbf{Chance} & 83.9\,\% & 78.9\,\% & 74.3\,\% & 69.9\,\% & 65.9\,\% \\
    \end{tabular}
  \label{table-fits-chance}
\vspace{-3mm}
\end{table}

\vspace{-2mm} 
\paragraph{\textbf{Phase 2: Finding (Gapless) Gadgets}}

In our experiments, we used Ropper~\cite{ropper} to locate gadgets in the
original binary and dynamically generate a ROP chain.
We chose a ROP chain with eight gadgets invoking \texttt{VirtualAlloc()} to
facilitate further code injection.
Note that the gadget locations are for the original binary,
but are unknown for the randomized process.

Thus, these gadgets are \textit{target code} in the notation of our
path-guided code re-discovery algorithm from Listing~\ref{listing-path-guided},
which the attacker can then simply invoke
to infer the addresses of the gadgets in the randomized process.
In the large binaries \texttt{xul.dll} and \texttt{mshtml.dll},
we had no problems to construct a ROP chain
from the gadgets reachable from found code anchors.

\begin{table}[H]
\centering
\vspace{-3mm}
\caption{Example for gapless gadgets.
The contained gadget is \texttt{pop ebx; ret} (5b c3) in all three cases.
Bytes encoding a register are bold, while gadgets are boxed.}
\vspace{-2mm}
\begin{tabular}{l|l|c}
\textbf{Hexadecimal} & \textbf{Disassembly} & \textbf{Gapless}\\
\hline
\hspace{7.8mm}\fbox{\textbf{5b}\hspace{0.1mm} c3} & \texttt{pop} ebx; \texttt{ret} & \xmark \\
\hspace{4.1mm}31 \fbox{\textbf{5b}\hspace{-0.2mm} \textbf{c3}} & \texttt{xor} [ebx-0x3d], ebx & \xmark \\
\textbf{b8} 00 \fbox{5b\hspace{1.15mm}c3} 00 & \texttt{mov} eax, 0x005bc300 & $\checkmark$ \\
\end{tabular}
\label{table-gapless}
\vspace{-3mm}
\end{table}

In Section~\ref{high-level-attack}, we described the Gadget Setup the attacker
has to perform to ``fill the gaps'' in the gadgets she wants to use.
However, randomization schemes without
instruction-level randomization beside register randomization,
allow to omit this phase, if
the attacker uses only
\textit{gapless gadgets} (see Table~\ref{table-gapless} for an example).
We call
a gadget gapless if it does not contain bits that encode a source or destination register of the original (and correctly aligned) instruction.
Table~\ref{table-xul} shows that roughly 10\,\% of the gadgets are gapless,
which leaves ten thousands of gapless gadgets for the attacker to choose from.

\begin{table}
\vspace{2mm}
\centering
  \caption{Statistics for the main libraries of the web browsers.}
  \vspace{-2mm}
    \begin{tabular}{l|r|r}
& \textbf{xul.dll} & \textbf{mshtml.dll} \\
\hline
Number of basic blocks & 1,601,390  & 793,527 \\
Portion of data bytes & 1.62\% & 2.39\% \\
\hline 
Number of gadgets & 1,641,199 & 689,891\\
Portion of gapless gadgets & 11.24\% & 8.97\% \\
    \end{tabular}
  \label{table-xul}
\vspace{-4mm}
\end{table}

\vspace{-2mm} 
\paragraph{\textbf{Phase 3: (Optional) Gadget Setup}}
The values in the gaps (the used
registers) can be derived by looking at the data flow.
From the original
binary, she knows which values flow into which instructions.  Thus, she
can usually infer the used registers, and thereby the gaps,
used in the next instructions from the ones she observes to be used upstream.
However, due to the large attack surface in the considered complex binaries,
we did not implement this step.

\vspace{-2mm} 
\paragraph{\textbf{Phase 4: Execution}}

We constructed a ROP chain for the randomized process,
but we still have to show that no gadget was garbled by DCR.
Discovering code anchors demands only a few disclosed bytes,
the code
re-discovery algorithm
destroys only the terminators which occur on a path between the code anchor and the
target gadget,
and the gadget setup destroys only code preceding the target gadget.
Thus, the target gadgets
are still intact,
and allow us to exploit the program.

\section{DCR Bypass of Randomized Instruc-{\newline}tions and Parameter Sequences}
\label{approach-level-2}

In addition to shuffled basic block locations and registers,
we now consider stronger randomization schemes which invasively transform
the code in each basic block beyond (syntactic) recognition,
e.\,g., insert NOPs, substitute instructions for equivalent ones, or
change the sequence of instructions.
This effectively thwarts locating a single gadget with certainty,
except probably for certain edge cases.
One could, however, use more coarse-grained code
fragments as gadgets: whole basic blocks or even whole functions.
As the latter would reduce attacks basically to a return-to-libc~\cite{ret2libc}
attack, we additionally
allow the randomization
scheme to change the sequence of parameters for functions at load time
to harden the system against function-reuse attacks.
In practice, such \textit{randomized parameter sequences} would be restricted to
non-exported functions and would be hard to implement reliably for binaries.

In contrast to our first attack, this attack thwarts stronger code randomization,
but at the price of an additional assumption: It assumes that a sufficient
number of function pointers can be leaked.
In practice however, this assumption is not hard to satisfy.
Large C\verb-++- applications provide hundreds of function pointers
through virtual function tables and it has been shown that other applications
also make heavy use of function pointers, e.\,g., as callbacks~\cite{control-jujutsu}.

\vspace{-2mm} 
\paragraph{\textbf{Phase 1: Finding Code Anchors}}
\label{approach-level-2-code-anchors}
In this attack, we leak
a number of function pointers to serve as code anchors.
To do so,
we leak an object that is located at a fixed offset
relative to the initial memory error and
use the object's virtual function table pointer to compute the
module's base address.
Ultimately, this allows us to harvest all virtual function tables~\cite{coop} to gain many code anchors.

\vspace{-2mm} 
\paragraph{\textbf{Phase 2: Finding Call Sites}}
The code randomization must preserve a function's semantics.
However, while we can use whole functions as gadgets, 
knowing a function's location is not sufficient to invoke it
due to the randomized parameter sequences.
We tackle this problem by leveraging our path-guided code re-discovery algorithm
to find a call site for each gadget function we aim to reuse.

\vspace{-2mm} 
\paragraph{\textbf{Phase 3: Reverse Engineering Parameter Sequences}}
Next, we read and disassemble the call sites
of our gadget functions.
Utilizing knowledge about the data flow in the call site's function from
the original binary,
we determine the sequence of the arguments at the call site.
We deem an argument's position to be deducible if we can
trace it to a data source which does not originate from the parameters of the
call site's function as those were randomized as well.
Effectively, this allows to reverse engineer the gadget function's randomized
parameter sequence.

Our function chain consists of \texttt{nsProcess} and \texttt{RunProcess}.
The former constructs the object and has no parameters,
the latter creates a process.
The crucial argument for \texttt{RunProcess} is \texttt{my\_argv},
which contains the command line of the to-be-created process.
Its first array element is initialized with the return
value of the function \texttt{ToNewUTF8String}.
Thus,
we can deduce the position of \texttt{my\_argv} by backtracing the data-flow
from the call-site \texttt{RunProcess} to the call-site of \texttt{ToNewUTF8String}.

\vspace{-2mm} 
\paragraph{\textbf{Phase 4: Execution}}

We gathered code anchors by
leaking function pointers without reading code memory.
Since we analyzed the call site functions, but not the gadget
functions, we did not read any memory we plan to execute.
According to Appendix~\ref{evaluation-code-usage}, there is only little code
executed by the process, so we can easily choose call site functions
the process does not execute.
Thus, our code-reuse chain is not garbled by DCR and we can successfully
exploit the process.

%% file: sections/defense.tex
\section{\tool{}: Byte-Granular DCR and XOM}
\label{defense}

\begin{figure*}
\centering
\includegraphics[width=0.95\textwidth]{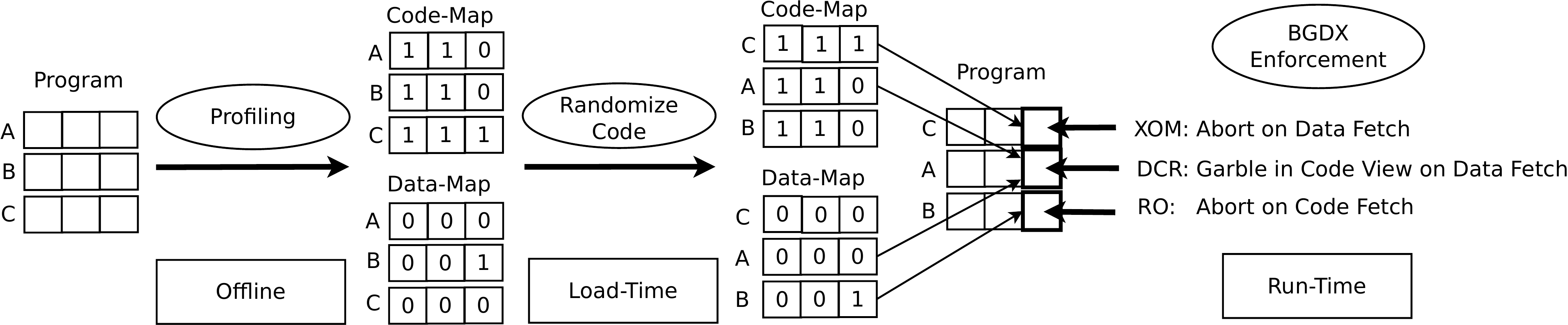}

\caption{Workflow of \tool{}, our proposed scheme to prevent JIT-ROP and code inference attacks.}
\label{figure-defense-high-level}
\end{figure*}

Execute-only memory (XOM) has become an important primitive for defensive research
and is vital for the resilience of, e.\,g., Readactor~\cite{readactor,readactor-plus-plus}.
While XOM has gained adoption for source code-based solutions, it
remains a challenge to transfer it
to binary-compatible schemes.
Destructive code reads offer binary-compatibility, but,
as we have shown, cannot fully prevent code inference attacks.
We now present \tool{} (Byte-Granular DCR and XOM),
our defensive scheme that leverages byte-granular
memory permissions to protect legacy binaries against code inference attacks.
Figure~\ref{figure-defense-high-level} shows \tool{}'s three main components:
first, the \textit{Profiling}
performs static and optionally dynamic analysis on the to-be-protected binary
to determine which bytes in executable memory are code, data or uncertain.
Identified code and data bytes are marked in the code-bitmap and data-bitmap,
respectively. Uncertain bytes are protected by a DCR policy,
so that the profiler can operate conservatively and thereby avoid
incorrectly identified bytes. Note that a relatively small percentage of
identified code bytes is sufficient for \tool{} to be effective and the
threshold can easily be reached through conservative static analysis or dynamic
analysis, as we show in Section~\ref{def-eval-security}.
Second, \textit{Code Randomization}
eliminates static relative offsets
and randomizes parameter sequences of internally used functions to hinder whole-function reuse.
Finally,
\textit{Memory Permission Enforcement} ensures that each byte in the code
section is protected in accordance with the profiling results. The byte
granularity is preserved regardless of the placement of identified code and data
bytes as well as unidentified bytes.

\subsection{Profiling}

For our static profiler component,
we extended the IDA Pro
disassembler~\cite{idapro} with heuristics to increase code coverage and
exclude identified code that has a low confidence.
First, we parse the executable file format and
harvest code pointers
to serve as starting
points for the subsequent code discovery.
Indirect control-flow transfers are resolved conservatively to avoid
false positives during code discovery. The next step iterates over discovered
code and tries to compute the address of as many memory reads as
possible. Bytes in code sections that are targeted by memory reads are marked
as data. Discovered code is marked as code. The output of the
static analysis consists of two bitmaps for each executable section of the binary,
one bitmap for the discovered code and the other for discovered data.
The static analysis is optimized for high confidence of recovered code and
data bytes rather than coverage. We wanted to avoid falsely identified code
bytes at all costs, because they may cause legitimate memory reads to crash.
Similarly, incorrectly recovered data bytes may cause legitimate code
execution to crash the program. For the general case static analysis is not
perfect and may introduce false positives. In
Section~\ref{def-eval-security} we show that the degree of code coverage
achieved with static analysis alone is sufficient to thwart code inference
attacks.

Our dynamic profiler component is based on the instrumentation framework
DynamoRIO~\cite{dynamorio}. During profiling, we record all instructions and
mark the instruction bytes as code. Furthermore, we intercept each memory
read and use its source address to resolve the corresponding module and
section. Memory reads that do not target any executable section of the
examined binary are ignored. Bytes in code sections that are targeted by
memory reads are marked as data bytes. The binary can be executed on
benchmarks or test benches during profiling in order to increase coverage.

The selection of profiling methods can be adjusted in accordance with the
requirements of the present use case. Static analysis produces good code
coverage, but detects only few data bytes.
Further, it cannot be completely ruled out that careful and conservative
static analysis introduces false positives.
In contrast, dynamic analysis has less code coverage,
but can detect more data reads and introduces no false positives.
Dynamic profiling allows to create perfect, although not complete, bitmaps.
Thus, it is suited to setup \tool{} completely fail-safe for COTS binaries.
Static and dynamic analysis results of the same binary can easily be merged
to complement one another. Conflicts should occur rarely and can be resolved
by marking the byte in question as uncertain.
In the end, the profiler incorporates the profiling results into the binary file's
executable file format.

\subsection{Code Randomization}

Similar to existing DCR schemes~\cite{near,heisenbyte},
we assume a code randomization component that
randomizes the code of the binary each time it is loaded and
adapts the profiling data accordingly.
We assume the following
three properties: first, the location of all basic blocks is randomized.
Second, the instruction locations within a basic block are randomized.
Third, the parameter sequences of internally used functions are randomized.

\subsection{Memory Permission Enforcement}

Our memory permission enforcement component supports the following three
policies that can be applied on a \emph{byte-granular} level: (1) read-only,
(2) execute-only and (3) destructive code read.

The profiling data
is utilized to protect each byte
of the code sections accordingly to Table~\ref{tab::policies}. Discovered code
is protected execute-only, which is also the default memory permission for code
pages. Thus, instruction fetches are not interrupted, which is the key factor
for efficiency. Data fetches to code pages are interrupted and
\tool{} looks up the permissions of the byte(s). If an execute-only byte is fetched
as data \tool{} immediately terminates the program.

To achieve byte granularity for read-only and DCR policies \tool{}
duplicates the physical pages containing such bytes, essentially creating code
and data views. Identified data bytes are garbled in the code view at
load-time to prevent their execution. However, data fetches to
data bytes are permitted and accomplished by redirection to the data view.
Uncertain bytes are protected by the DCR policy which means that they are
initially intact in both views. Data fetches to DCR bytes succeed analogous to
read-only bytes, but additionally the corresponding bytes are garbled in the code
view on-the-fly in order to prevent their subsequent execution. DCR bytes can
be executed analogous to execute-only bytes, until they are garbled in the code view
by data fetches.
Section~\ref{implementation} gives further
implementation details of our \tool{} prototype.

In terms of security, there are two key differences between \tool{} and the
original DCR implementations~\cite{near,heisenbyte}.
First, the attacker cannot execute
legitimate data in code memory, such as jump tables, if they have been identified
during profiling. Data may contain
exotic and lucrative gadgets due to their different nature than code, so
\tool{} further reduces the attack surface. However, the second effect is much more
important, as it severely impedes code inference attacks: random illegitimate
memory reads targeting the code section have an inherent probability to hit
execute-only protected bytes, which results in immediate program termination.
Since code inference attacks need many such probes, this significantly reduces
the probability of successful exploitation, as we will show in
Section~\ref{def-eval-security}. An attacker may try to leverage
code pointers to avoid blind probing. However, \tool{} also leverages all
accessible code pointers during profiling, so that code reachable through code
pointers is protected execute-only and thereby useless for code inference.

%% file: sections/implementation.tex
\section{Implementation Details}
\label{implementation}

In this section, we cover implementation details of our prototype
implementation of \tool{}.
Our memory permission enforcement component leverages existing virtualization technology.
Hypervisors rely on SLAT (Second Level Address Translation) in order to map physical pages of the
guest system to physical pages of the host system.
Intel's hardware-assisted implementation of SLAT is called EPT (Extended Page Tables),
which we leverage
to enforce the different byte-granular memory
permissions shown in Table~\ref{tab::policies}.

\begin{table}
\centering
\caption{Enforced policies, their encoding in the profiling metadata and the system's actions when instruction and data fetches occur.}
\vspace{-2mm}
\small
\setlength\tabcolsep{1.3mm}
\begin{tabular}{c|c|c|c|c|c}
\textbf{Policy} & \multicolumn{2}{c|}{\textbf{Profiling Metadata}} & \textbf{Meaning} & \textbf{Event} & \textbf{Action} \\
 & \textbf{Code-Bit} & \textbf{Data-Bit} & & & \\
\hline
XOM & 1 & 0 & Code-Byte & IF\textsuperscript{1} & \checkmark \\
 & & & & DF\textsuperscript{2} & \xmark \\
\hline
RO & 0 & 1 & Data-Byte & IF\textsuperscript{1} & \xmark \\
 & & & & DF\textsuperscript{2} & \checkmark \\
\hline
DCR & 0 & 0 & Uncertain & IF\textsuperscript{1} & \checkmark/\xmark\\
 & & & & DF\textsuperscript{2} & \checkmark\\
\multicolumn{6}{l}{Instruction fetch\textsuperscript{1}, Data fetch\textsuperscript{2}}
\end{tabular}
\label{tab::policies}
\vspace{-4mm}
\end{table}

Normally,
the EPT contains a 1:1 mapping of guest physical pages to host
physical pages.
Windows On-Demand Paging causes virtual pages to be mapped only
after they are touched the first time. This initial page access causes a page-not-present fault which is handled
by the kernel's page fault handler. The memory management then maps the appropriate physical page at the
faulting address. We augment the page fault handler to call into the hypervisor
to create a shadow page and point the EPT entry of the guest physical page to the host physical address of
the shadow page.
Additionally, we set the permissions of the shadow page to execute-only
in the EPT entry. Essentially, this creates separate data and code views for the same page,
as used by the DCR implementations
(see Figure~\ref{fig:DCR-ROP} in Section~\ref{background}).
During on-demand paging, our hypervisor
overwrites the code view of bytes marked as read-only in the profiling metadata
with \texttt{INT 3} instructions,
which enforces the read-only policy by preventing the bytes from being executed.
Note that the original bytes are still intact in the data view.
Due to the EPT setup and the permissions of the code view, instruction fetches do not raise page faults or EPT faults,
which is a key element for efficiency.

However, data fetches \textit{do} cause EPT faults, which direct control to our
modified page fault handler in the hypervisor.
There, we disassemble the faulting instruction to get the operand
size and the number of bytes of the attempted read.
The profiling metadata of each byte is checked in order to detect attempts to read
a code byte, as such reads violate the XOM policy and must therefore result in process termination.
Bytes marked as data or uncertain are allowed to be read.
Otherwise, we point the EPT entry to the data view to
enable legitimate data fetches.
Additionally, if a read byte is marked as neither code nor data,
the DCR policy is applied, and
we overwrite the byte with \texttt{INT 3} instructions in the code view.
Lastly,
by utilizing the single-step flag,
we redirect the EPT entry to the execute-only view to facilitate
fast instruction fetches
before returning control to the instruction which caused the data fetch.

%% file: sections/evaluation.tex
\section{Evaluation}

In this section, we evaluate our prototype implementation of \tool{}
with regards to performance and security.
All experiments were executed inside a
virtual machine running Windows~7 on an Intel Core i7-2640M
@ 2.8GHz with 8~GB DDR3-RAM @ 1600MHz.

\subsection{Benchmarks}
\label{eval-benchmarks}

First, we evaluate the performance overhead
introduced
by our defense system.
To this end, we used the SPEC CPU 2006 Integer Benchmark Suite~\cite{spec2006}.

However, two of the benchmarks (\texttt{403.gcc} and \texttt{462.libquantum})
did not compile using the default configuration.
The first one uses variables beginning with two underscores,
which is reserved for compiler use in MSVC.\footnote{https://msdn.microsoft.com/en-us/library/2e6a4at9.aspx}
The second one requires complex arithmetic functions,
which are not supported by MSVC,
despite the fact that they are specified in the C99-standard.
The SPEC2006-FAQ explicitly mentions that this benchmark is problematic.\footnote{https://www.spec.org/cpu2006/Docs/faq.html}
However, the SPEC2006 System Requirements explicitly states in section I.A.2
that other compilers, like MinGW, are not recommended.\footnote{https://www.spec.org/cpu2006/Docs/system-requirements.html}
A third benchmark (\texttt{400.perlbench}) does compile,
but attempts to access member fields which are internal on
Windows, and therefore crashes when executed on the supplied test inputs.
Again, this is a known problem,\footnote{https://www.spec.org/cpu2006/src.alt/}
and while there is a patch available,
it refuses to work on the specific version of SPEC available to us.
Note that these three problems are not caused by our system,
but by the Windows environment which
is only partially supported by SPEC.

On the remaining nine benchmarks, we achieved a
geometric mean overhead of 3.95\,\%
with a median of 4.53\%
and a worst-case of 7.21\%
(see Table~\ref{tab::spec}).
These benchmarks are computationally intensive,
and we therefore
deem our system to induce a minimal overhead in practice.

\begin{table}
\centering
\caption{Runtime overhead for the SPEC 2006
Benchmarks.
Benchmarks marked with \dag{} are not supported on Windows.}
\vspace{-2mm}
\setlength{\tabcolsep}{1.5mm}
\small
\begin{tabular}{l|c||l|c}
\textbf{Benchmark} & \textbf{Overhead} & \textbf{Benchmark} & \textbf{Overhead}\\
\hline
400.perlbench  & \dag       & 458.sjeng      & ~5.58\,\%\\
401.bzip2      & ~4.53\,\% & 462.libquantum & \dag\\
403.gcc        & \dag & 464.h264ref    & ~4.53\,\%\\
429.mcf        & ~7.21\,\% & 471.omnetpp    & ~0.76\,\%\\
445.gobmk      & ~6.25\,\% & 473.astar      & ~4.06\,\%\\
456.hmmer      & ~3.75\,\% & 999.specrand   & ~3.85\,\%\\

\end{tabular}
\label{tab::spec}
\vspace{-3mm}
\end{table}

\subsection{Firefox Browser Benchmarks}

In addition,
we applied our system to the Firefox
browser and used popular browser benchmarks to demonstrate a more
realistic workload. In particular, we used Speedometer~\cite{speedometer},
Peacekeeper~\cite{peacekeeper}, and JetStream~\cite{jetstream}:

\begin{itemize}
\item The Speedometer benchmark provides the number of repetitions per minute for
typical browser interactions powered by different JavaScript frameworks.
While the baseline Firefox performed, on average (arithmetic mean)
28.6 runs/min, the secured version performed only 26.1 runs/min.
This corresponds to an overhead of 9.57\,\%.
\item The JetStream benchmark performs a series of computation heavy operations,
as they would occur in different phases of feature-rich web-applications.
For each benchmark, it reports so-called ``scores'',
where higher scores are better.
Ultimately, it calculates an overall-score using the geometric mean.
The baseline Firefox had a score of
105.57, while JetStream reported a score of 88.319 for the secured version.
Since these scores loosely correspond to consumed time,
we estimate an overhead of 16.34\,\%.
\item The Peacekeeper benchmark also aggregate a series of smaller benchmarks.
However, only the ones performing computer graphic computations and reporting
Frames per Second (FPS) have a linear relation to consumed time.
Those average to an overhead of 3.16\,\%.
\end{itemize}

Given that every benchmark-suite has its own scoring system,
it is hard to give an overall result.
Our best estimation is an aggregation to a
geometric mean overhead of 7.9\,\%,
with a median of 9.57\,\%.
Nevertheless, these benchmarks confirm that \tool{}'s overhead is moderate
in practice for complex, real-world applications such as a web browser.

\subsection{Security Considerations}
\label{def-eval-security}

To bypass DCR, the attacker exploits that code memory is readable to infer the 
location of gadgets.
In contrast to Heisenbyte~\cite{heisenbyte},
every byte confirmed as code reduces \tool{}'s attack surface,
since
a read attempt of confirmed code immediately crashes the application.

\vspace{-2mm}
\paragraph{\textbf{Memory probing.}}
The attacker is reduced to random probing for two reasons:
first, because the load-time code randomization invalidates prior
knowledge about code locations.
Second, because traversing the control-flow graph is mostly prohibited,
since
easily traceable control-flow transitions also increase the defender's
code-coverage, which in turn increases the ratio of XOM-protected bytes.

\vspace{-2mm}
\paragraph{\textbf{Successful attack.}}
\label{eval-succ-attack}
Given that an imperfect code coverage results in a probabilistic defense,
we deem an attack to be successful, if enough gadgets for
a ROP chain can be gathered at least once in a billion tries,
i.\,e., the attacker's chance of success is at least $10^{-12}$.\\

To evaluate \tool{}'s security, we have to estimate the number of
necessary random probing attempts.
Current ROP defenses\cite{ropecker,kbouncer} assume that a ROP chain needs at
least eight gadgets, but
a recent ROP defense bypass~\cite{len-rop-chain}
took special care to construct a ROP chain with only five gadgets.
Thus, we assume here that a ROP chain needs only five gadgets.
Gadgets are divided into classes~\cite{heisenbyte-18} to distinguish
their different tasks.
However, these classes are not distributed equally, and the attacker needs gadgets
from different classes.
A Monte-Carlo simulation using a published gadget distribution~\cite{heisenbyte-18}
revealed that an attacker needs to find on average 37.2 random gadgets to gather one of each
necessary class.
Taking into account that code randomization leaves on average only about 33\,\%
of the gadgets in a class intact~\cite{heisenbyte-18}, the number of tries has to be multiplied by 3.3.
Table~\ref{table-xul} shows that there is roughly one gadget per basic block
and we have shown
in Section~\ref{evaluation-finding-code-anchors}
that an attacker needs about 3.3 random probes to find a code anchor
at basic block level.
To favor the attacker's odds, we assume that a code anchor immediately identifies
the gadget in the basic block without further probes or gadget setup.
However, taking the basic blocks' sizes into account
(see Appendix~\ref{appendix-bb-sizes}),
roughly 57\,\% of the gadgets are destroyed by probing,
assuming that a gadget has at least two bytes and that a probe destroys
at least three bytes.
Thus, the number of necessary tries has to be multiplied by 2.3.
Due to the shuffling of basic blocks, the execute-only protected code as well as the
gadgets are close to uniformly distributed in the process memory.
Thus, a conservative estimation of the number of random probing attempts yields:
$37 \cdot 3 \cdot 3 \cdot 2 = 666$

\begin{figure}[h]
\centering
  \includegraphics[width=0.48\textwidth]{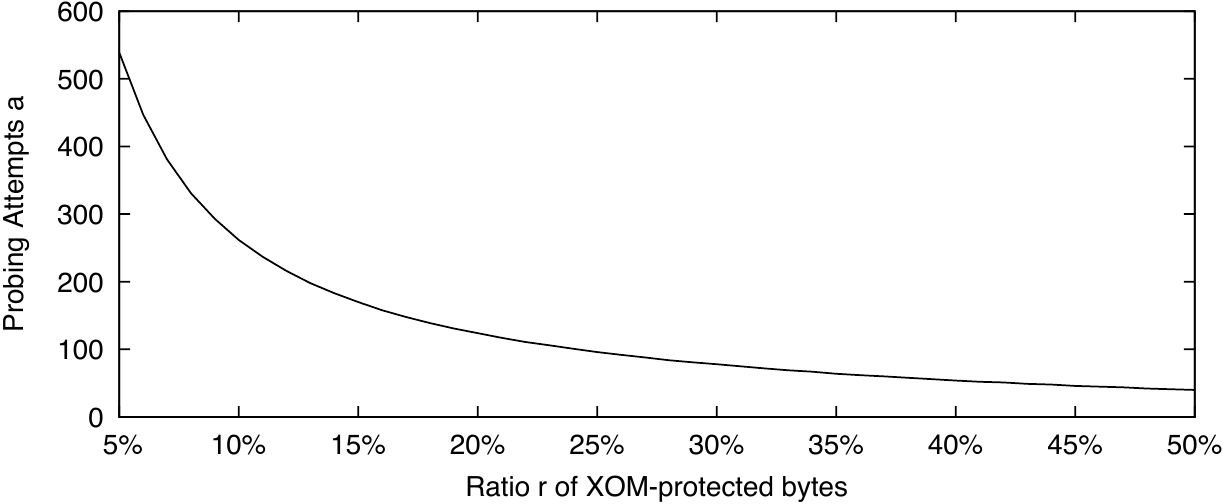}
\vspace{-5mm}
  \caption{Number of probing attempts $a$ the attacker can make before his success probability sinks to $10^{-12}$, given the ratio $r$ of XOM-protected bytes in code memory.
The graph uses the formula ${(1 - p)}^{a} = 10^{-12}$.}
  \label{figure-success-prob}
\end{figure}

This results in a ${(1-p)}^{666}$ chance of success, with $p$ being the ratio
of bytes protected by XOM.
Figure~\ref{figure-success-prob} shows that the ratio of a priori identified code
bytes can be as low as 5\,\%
and still prevent the attacker from carrying out the necessary probes.
That is, even a 5\,\% code coverage reduces the attacker's chance of a successful
code inference attack to less than $10^{-12}$.

\vspace{-2mm}
\paragraph{\textbf{Achievable code coverage.}}
\label{sec-browser-code-coverage}
To estimate realistically achievable code coverage, we executed the JetStream~\cite{jetstream}
and Peacekeeper~\cite{peacekeeper} browser benchmarks.
Together, they executed 540,068 of 1,601,390 basic blocks for Firefox's \texttt{xul.dll},
which corresponds to roughly 33\,\% code coverage. This analysis took only
a few minutes and is easily repeatable.
However, given that there is debug information available
for both Firefox and Internet Explorer,
their static analysis leads to almost perfect code coverage.
Table~\ref{tab-spec-profile} also shows that executing even the
small test sets accompanying the SPEC benchmark can achieve a substantial
code coverage.

\vspace{-2mm}
\paragraph{\textbf{Summary.}}
We have shown that the practically achievable code coverage on binaries is
sufficient to limit the number of probing attempts to a level,
which reduces the chances of a code inference attack to practically zero (see Section~\ref{eval-succ-attack}).
Thus, unlike DCR, \tool{} prevents sophisticated code inference attacks.
Together with randomized parameter sequences, it can further prevent
whole-function ROP.
Since it imposes only a small overhead (see Section~\ref{eval-benchmarks}),
does not demand to perfectly separate code from data,
and can protect binaries without access to source code or debug information,
it can efficiently protect legacy binaries.
Table~\ref{table:dcr-xom-hybrid} summarizes these results, showing that \tool{}
unites the security benefits of XOM with the legacy
binary compatibility of DCR.

\vspace{-5mm}

\begin{table}[h]
\centering
\caption{Bytes confirmed as code, respectively data, through dynamic analysis by executing the SPEC benchmark applications on their test sets.
For Firefox's and Internet Explorer's main-libraries, Appendix~\ref{evaluation-code-usage}
shows that roughly 10-15\,\% code coverage is achievable by visiting 100 popular websites,
whereas Section~\ref{sec-browser-code-coverage} shows that roughly 33\,\% can be achieved
by executing popular browser benchmarks.
However, since debug information is available for both browsers,
Table~\ref{table-xul} can give accurate numbers for both code and data.}
\vspace{-2mm}
\small
\begin{tabular}{l|r|r|r}
\textbf{Benchmark} & \textbf{Bytes in} & \multicolumn{2}{r}{\textbf{Confirmed to be}} \\
                   & \textbf{text section} & \textbf{Code} & \textbf{Data}\\
\hline
401.bzip2      & 93,600 & 49.54\,\% & 0.10\,\% \\
429.mcf        & 66,184 & 36.86\,\% & 0.06\,\% \\
445.gobmk      & 603,648 & 29.77\,\% & 0.77\,\% \\
456.hmmer      & 201,208 & 24.71\,\% & 0.05\,\% \\
458.sjeng      & 140,320 & 36.31\,\% & 0.26\,\% \\
464.h264ref    & 463,672 & 29.94\,\% & 0.05\,\% \\
471.omnetpp    & 606,632 & 20.45\,\% & 0.03\,\% \\
473.astar      & 94,160 & 47.27\,\% & 0.07\,\% \\
999.specrand   & 46,192 & 31.37\,\% & 0.15\,\% \\

\end{tabular}
\label{tab-spec-profile}
\end{table}

\begin{table}[t]
\centering
  \caption{Comparison of DCR, XOM and \tool{}.}
\vspace{-1mm}
\small

\setlength\tabcolsep{2mm}
    \begin{tabular}{l|c|c|c}
 & \textbf{DCR} & \textbf{XOM} & \textbf{\tool{}} \\
\hline
\textbf{Secure against} & & & \\
~Conventional JIT-ROP & \checkmark & \checkmark & \checkmark \\
~Code inference attacks & \xmark & \checkmark & \checkmark \\
~Whole-function ROP & \xmark & \xmark & \checkmark \\
\textbf{Qualification criteria} & & & \\
~Data reads from code memory & \checkmark & \xmark & \checkmark \\
~Legacy binary support & \checkmark & \xmark & \checkmark \\

~Efficient & \checkmark & \checkmark & \checkmark \\
    \end{tabular}
  \label{table:dcr-xom-hybrid}
\vspace{-3mm}
\end{table}

%% file: sections/discussion.tex
\section{Discussion}
\label{discussion}

We now discuss our results and open challenges.

\paragraph{\textbf{Simulated attack environment.}}

There is no implementation of a DCR system with a fine-grained code
randomization publicly available, so we had to simulate both to test our attacks.
As for DCR,
logging data and instruction fetches allows to
precisely evaluate whether exploitation has succeeded.
As for code randomization,
our memory probing abstraction layer
provides the same information to the attacker.
Thus, our attacks demonstrate that DCR can be bypassed
even with substantially stronger code randomization
than required by DCR or publicly available for legacy binaries.

\paragraph{\textbf{Defense limitations.}}

Our defense scheme generates memory permission bitmaps to (imperfectly)
distinguish code and data.
Since these bitmaps are created before code randomization takes place,
they could be invalid at runtime.
Thus, we require an interplay between code randomization and our memory
permission bitmaps,
namely that the code randomization updates the bitmaps when performing its transformations.
To the best of our knowledge, any
randomization technique
could be adapted to this requirement.

To hinder whole-function reuse attacks, we randomized the
parameter sequence of functions.
This
is no problem for internally used functions,
but the parameter sequence of exported functions
cannot be randomized, because all processes use
the same parameter sequence to invoke the function.
Thus, exported functions are at risk for whole-function reuse.
To tackle this problem,
one could provide
separate instances of the shared library for each process.

Lastly, similar to other binary-based security frameworks~\cite{cfi-journal,near,heisenbyte-27}, 
we do not specifically handle just-in-time (JIT) compiled code.
However, it is easily possible to integrate the JIT-protection techniques in \tool{}:
We can use DCR, just as the Heisenbyte system does~\cite{heisenbyte}.
If the JIT-Compiler would provide code and data bitmaps, we could also provide the
stronger protection offered by our XOM and read-only policies.

\paragraph{\textbf{Separating Code and Data.}}
Separating code and data is an important step for disassembly and therefore
also for defense schemes working on binaries.
Recent work by Andriesse et al.~\cite{andriesse2016dissassembly} has shown that intermingled
code and data is, in contrast to the popular opinion,
\textit{not} present in pure-C applications on x86/64,
if they are compiled with modern \texttt{GCC} or \texttt{Clang} compilers.
However, they also state that it does frequently occur on Windows,
due to the \texttt{MSVC} compiler.
It should be noted that the library code is often optimized and therefore
contains hand-written assembly.
Unless the programmers take special care to write separation-friendly
code, such code contains intermingled code and data, despite the compiler's effort.
E.\,g., the \texttt{glibc} library, which is present in roughly 97\,\% of about
2000 ELF-binaries on a commodity Linux system, grew increasingly disassembler-friendly
over the last versions.
As reported by Meng and Miller~\cite{binary-code-is-not-easy},
\texttt{glibc} contained intermingled code and data as recently as 2016.
Furthermore, Andriesse et al. state in a blog post~\cite{andriesse-blog-post}
that both \texttt{GCC} and \texttt{Clang} do not separate code and data
on ARM platforms.

With the widespread use of ARM and \texttt{glibc},
combined with the fact that an ELF file links, on average, against 10.2 libraries,
each of which may contain hand-written assembly,
we think it is fair to say that separating code from data is still a problem
in context of legacy binaries.
In addition, Table~\ref{table-xul} shows that roughly 2\,\% of the code section in
browsers' main libraries are data bytes---if it were not for the provided debug
information, supporting such complex binaries would be a struggle for binary-only
defense schemes.

%% file: sections/related-work.tex
\section{Related Work}
\label{related-work}

We have already discussed various code-reuse attacks and defenses throughout this paper and hence center 
our discussion on recent code-reuse attack techniques in the context 
of DCR protection and closely related topics.

\paragraph{\textbf{Extended Code-Reuse Attacks}}
Counterfeit Object-Oriented Programming (COOP)~\cite{coop} describes a powerful attack against C\verb!++!
applications bypassing many current techniques for control-flow
integrity (CFI). It builds a chain of virtual methods to generate arbitrary malicious computation. 
Crash resistance~\cite{crash-resistance}
allows specific code constructs to probe arbitrary (even unmapped) memory to disclose 
randomized code and data sections without crashing the process. 
Lastly, the so-called indirect JIT-ROP 
attack techniques harvest code pointers from data sections to infer gadget locations~\cite{isomeron}. 
Only the first two techniques require reading gadgets before utilizing them,
but it is not clear, if these attacks can be adapted to randomized parameter sequences,
a challenge we successfully tackled in Section~\ref{approach-level-2}.

\paragraph{\textbf{Execute-only Memory}}
JIT-ROP attacks exploit the fact that code can be read as data.
As a consequence, the concept of execute-only memory (XOM) was introduced in a tool
called XnR~\cite{xnr}, which uses the page-fault handler to distinguish between
allowed code reads for purposes of execution and forbidden code reads, where
code is read as data. 
HideM~\cite{hidem} uses the translation lookaside buffer (TLB), which
has separate caches for code and data access, to make the same distinction.
Readactor~\cite{readactor} uses a hardware virtualization scheme
to make certain areas of the code memory non-readable. In addition, 
it also replaces each code pointer by a pointer to a trampoline,
thereby hiding the actual code pointers and preventing their leakage in indirect JIT-ROP attacks~\cite{isomeron}.
Readactor\verb!++!~\cite{readactor-plus-plus} enhances Readactor with regard to the
specifics of C{}\verb!++! to also account for COOP attacks.
Naturally, this pointer indirection increases the overhead in comparison to pure XOM.

However, XOM-based defenses suffer from a major shortcoming: they require perfect separation of code and
data, as data misclassified as code would lead to loopholes for the attacker
and code misclassified as data results in application crashes.
This makes it very hard to apply XOM for legacy binaries. In fact, 
XnR requires source code, Readactor additionally a modified compiler,
while HideM has to sacrifice security for compatibility.

Naturally,
every byte protected by DCR rather than XOM poses an opportunity for the attacker.
Nevertheless, our evaluation in Section~\ref{def-eval-security} demonstrates that 
\tool{} achieves almost the same level of security
on legacy binaries than XOM approaches with byte-granular
protection~\cite{readactor,readactor-plus-plus} achieve with source code.

\paragraph{\textbf{Information Hiding}}
Multiple approaches try to limit the attacker's capability to
locate usable code memory.
One example is Oxymoron~\cite{oxymoron},
which cuts code into single memory pages,
and tries to hide code pointers from one page to another.
ASLR-Guard~\cite{aslr-guard} combines classical shadow stacks with an isolated
memory region for code pointers to hide them.
Readactor~\cite{readactor} and its successor~\cite{readactor-plus-plus}
combine XOM with trampolines to hide code pointers. 
Code-pointer integrity hides code pointers in a dedicated safe memory region~\cite{cpi}.
Unlike our proposed defense system, all these techniques
require source code and thus cannot protect legacy binaries.

\paragraph{\textbf{Control-Flow Integrity (CFI)}}
Arguably, the problem is not in reading code memory,
but in the diversion from the program's planned control flow
to execute code of the attacker's choice.
Thus, control-flow integrity (CFI)
aims to prevent such diversion~\cite{cfi-journal}.
Modern approaches such as
binCFI~\cite{cfi-for-cots},
or CCFIR~\cite{ccfir}
support binaries as well. However, their security suffers from imprecise
analysis or coarse-grained policies~\cite{coop, out-of-control}.

\paragraph{\textbf{Live Randomization}}
One could also
prevent code inference attacks by repeatedly performing code randomization while the 
application executes~\cite{tanenbaum-rerandomization}.
This ensures that the
discovered gadget locations are outdated by the time the attacker actually invokes them. 
Modern approaches~\cite{timely-rerandomization, remix, shuffler} have
a low performance overhead, e.\,g., by randomizing only after events,
which could leak information.
Unlike our defense system, they require source code or debug symbols,
and therefore cannot protect legacy applications.

%% file: sections/conclusions.tex
\section{Conclusions}
\label{conclusions}

In this paper, we show generic attacks that bypass destructive code read
defenses, like Heisenbyte and NEAR.
More specifically, we constructed two different attacks
against recent versions of the popular Firefox web browser (32-bit) and
Internet Explorer (64-bit) protected by DCR.
In contrast to prior work, our attacks consider properly applied,
strong load-time code randomization and thus
demonstrate that DCR's fundamental assumption that one has to
read a gadget in order to execute it, does not hold in practice.

Furthermore, we propose a novel defensive scheme called \tool{} that applies memory
permissions on a byte-granular level.
\tool{} combines the security benefits of execute-only memory with the
legacy binary compatibility of previous DCR schemes.
This defense has only a small performance overhead and
needs only a manageable portion of code
marked as execute-only to render code inference attacks impractical.

In summary,
we demonstrate that schemes relying solely on DCR are conceptually broken,
and propose a novel defense
to protect legacy binaries against
code inference attacks.

%% file: sections/acknowledgements.tex
\section*{Acknowledgments}

We thank the anonymous reviewers and our shepherd Aravind Prakash for their
valuable feedback. This work was supported by the German Research Foundation
(DFG) research training group UbiCrypt (GRK 1817) and collaborative research center CROSSING (CRC 1119, Project S2), and by the European Research
Council (ERC) under the European Unions Horizon 2020 research and innovation
programme (ERC Starting Grant No. 640110 (BASTION)).

%% file: sections/appendix.tex
\appendix
\section{Basic Block Sizes}
\label{appendix-bb-sizes}

\begin{figure}[h]
\centering
  \includegraphics[width=0.49\textwidth]{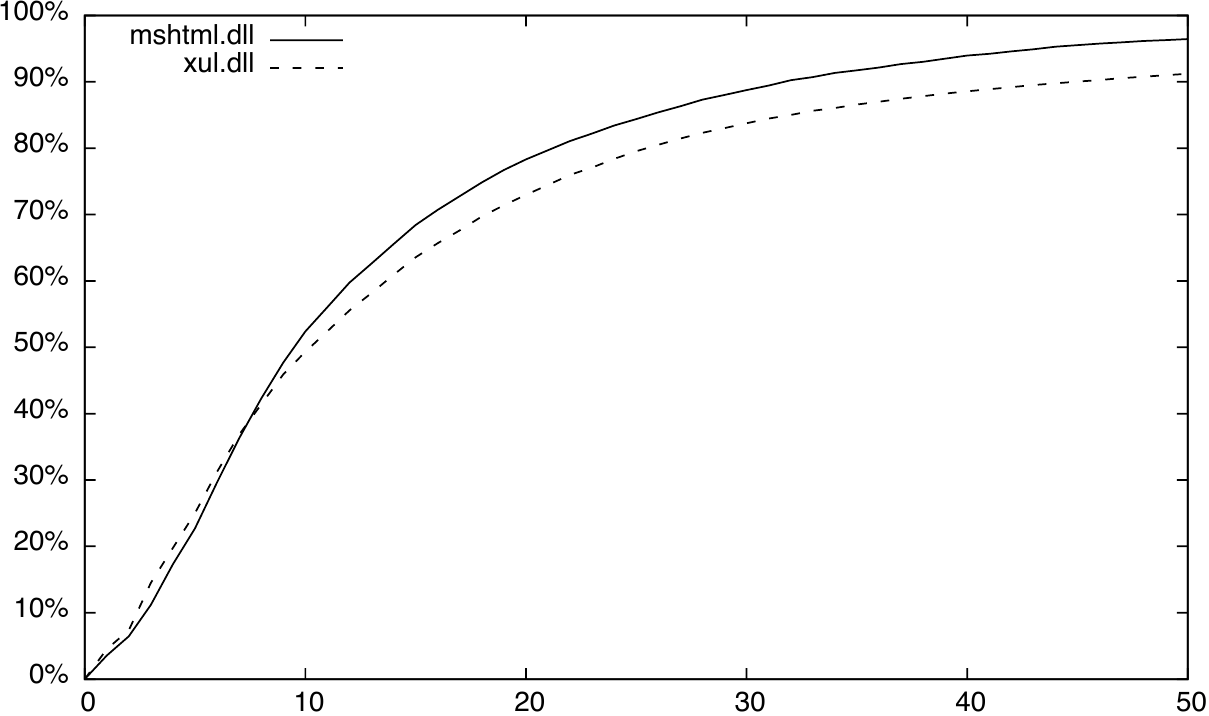}
  \caption{Cumulative density function of the basic block sizes in \texttt{xul.dll} (32-bit) and \texttt{mshtml.dll} (64-bit).}
  \label{figure-bb-sizes}
\end{figure}

Figure~\ref{figure-bb-sizes} shows the sizes of the basic blocks in the main libraries
\texttt{xul.dll} and \texttt{mshtml.dll} for the web browsers Firefox, respectively
Internet Explorer. In particular, it gives a cumulative density function,
i.\,e., it shows that roughly 20\,\%
of the basic blocks in either library are
no larger than five bytes, while basic blocks with ten bytes or less
comprise 50\,\% of the basic blocks.

\section{Portion of Executed Code}
\label{evaluation-code-usage}
The attacker may crash the application while probing, since she cannot predetermine
which code she probes and may therefore accidentally read memory
the application later executes.
To quantify this risk,
we used DynamoRIO~\cite{dynamorio} to count
how often the basic blocks of the two browsers' main
libraries are executed while visiting the Alexa Top-100 websites.
As Table~\ref{table-alexa} shows,
about 80\,\% of the libraries' code is not executed at all.
Taking into account that
(1) an attack takes place after loading a website when most of the code
has already been executed,
and (2) that the victim only needs to visit \textit{one} website instead of a hundred,
only about 1.27\,\%, respectively 4.46\,\%, of the code
poses a risk for the attacker.
Due to
the high number of gadgets (see Table~\ref{table-xul}),
there are high chances to find a gadget.
Thus,
there is a low risk for the attacker to crash the randomized application.

\begin{table}[H]
\centering
  \caption{Basic block execution frequencies for the browsers' main libraries when visiting the Alexa Top-100 websites. Basic block executions occurring during startup are subtracted, as they occur before any attack.}
\vspace{-2mm}
    \begin{tabular}{c|r|r}
\textbf{\#Executed} & \textbf{xul.dll (32-bit)} & \textbf{mshtml.dll (64-bit)} \\
\hline
$x = 0$ & 83.27\,\% & 83.77\,\% \\
$1 \le x < 100$ & 15.45\,\% & 11.75\,\% \\
$x \ge$ 100 & 1.27\,\% & 4.46\,\% \\
    \end{tabular}
  \label{table-alexa}
\end{table}